\documentclass[12pt,preprint]{aastex}

\usepackage{lscape}
\usepackage{natbib}

\usepackage{graphicx}

\shorttitle{Class I methanol maser in EGOs} \shortauthors{Xi Chen et
al.}

\begin{document}


\title {A 95 GHz Class I Methanol Maser Survey Toward GLIMPSE Extended Green Objects (EGOs)}
\author
 {Xi ~Chen\altaffilmark{1, 2}, Simon P. Ellingsen\altaffilmark{3}, Zhi-Qiang ~Shen\altaffilmark{1, 2}, Anita Titmarsh \altaffilmark{3},
 and Cong-Gui Gan\altaffilmark{1, 2}}

\altaffiltext{1} {Key Laboratory for Research in Galaxies and
Cosmology, Shanghai Astronomical Observatory, Chinese Academy of
Sciences, Shanghai 200030, China; chenxi@shao.ac.cn}

\altaffiltext{2} {Key Laboratory of Radio Astronomy, Chinese Academy
of Sciences, China}

\altaffiltext{3} {School of Mathematics and Physics, University of
Tasmania, Hobart, Tasmania, Australia}


\label{firstpage}


\begin{abstract}

We report the results of a systematic survey for 95 GHz class I
methanol masers towards a new sample of 192 massive young stellar
object (MYSO) candidates associated with ongoing outflows (known as
extended green objects or EGOs) identified from the \emph{Spitzer}
GLIMPSE survey. The observations were made with the Australia
Telescope National Facility (ATNF) Mopra 22-m radio telescope and
resulted in the detection of 105 new 95 GHz class I methanol masers.
For 92 of the sources our observations provide the first
identification of a class I maser transition associated with these
objects (i.e. they are new class I methanol maser sources). Our
survey proves that there is indeed a high detection rate (55\%) of
class I methanol masers towards EGOs. Comparison of the GLIMPSE
point sources associated with EGOs with and without class I methanol
maser detections shows they have similar mid-IR colors, with the
majority meeting the color selection criteria
-0.6$<$[5.8]-[8.0]$<$1.4 and 0.5$<$[3.6]-[4.5]$<$4.0. Investigations
of the IRAC and MIPS 24 $\mu$m colors and the associated millimeter
dust clump properties (mass and density) of the EGOs for the
sub-samples based on which class of methanol masers they are
associated with suggests that the stellar mass range associated with
class I methanol masers extends to lower masses than for class II
methanol masers, or alternatively class I methanol masers may be
associated with more than one evolutionary phase during the
formation of a high-mass star.

\end{abstract}

\keywords{masers -- stars:formation -- ISM: molecules -- radio
lines: ISM --
 infrared: ISM}

\section{Introduction}

Methanol masers are quite common in massive star forming regions
(MSFRs). Historically they have been empirically classified into two
categories (class I and class II). The initial classification was
based on the sources towards which the different transitions were
detected (Batrla et al. 1988; Menten 1991a). Class II methanol
masers are often found to be associated with strong centimeter
continuum emission within 1$''$ (e.g. ultracompact (UC) regions;
Walsh et al. 1998), infrared sources and OH masers. The strongest
and best known class II transition is the 5$_{1}-6_{0}$ A$^{+}$ line
at 6.7 GHz. Since its discovery (Menten 1991b), this class bright
maser has become a reliable tool for detecting and studying regions
where massive stars form and are in their very early stages of
evolution (e.g. Minier et al 2003; Bourke et al 2005; Ellingsen
2006; Xu et al 2008; Pandian et al 2008). A number of surveys have
been conducted for the 6.7 GHz class II methanol maser, resulting in
the detection of $\sim$ 900 sources in the Galaxy to date. The
surveys include those that are summarized in the compilation of
Pestalozzi et al. (2005) and the recent searches of Pandian et al.
(2007), Ellingsen (2007), Xu et al. (2008; 2009) and the Parkes
Methanol Multibeam (MMB) blind survey ($\sim$ 300 sources (Green et
al. 2009), but so far only part published by Caswell et al. 2010 and
Green et al. 2010). In contrast class I methanol masers are usually
found offset by 0.1 - 1.0 pc from UC H{\sc ii} regions, infrared
sources and OH masers (e.g. Plambeck \& Menten 1990; Kurtz et al.
2004). A number of observations of class I methanol masers showed
that they were located at the interface between molecular outflows
and the parent cloud (Plambeck \& Menten 1990; Johnston et al. 1997;
Kurtz et al. 2004; Voronkov et al. 2006). These early results
suggested that class I and class II methanol masers favored very
different environments. These observational findings were supported
by early theoretical models of methanol masers which found that
class I masers are collisionally excited, in contrast to class II
masers which have a radiative pumping mechanism (Cragg et al. 1992).
Observations made since the mid-1990s (e.g. Slysh et al. 1994) have
shown that at single dish resolutions class I and class II masers
are often associated. Observations at high spatial resolution (e.g.
Cyganowski et al. 2009) show that while the two types of masers are
typically not co-spatial on arcsecond scales, they are usually
driven by the same young stellar object.

Compared to class II masers, class I methanol masers are relatively
poorly studied and understood. There have only been a few large
surveys for class I masers (mainly at 44 and 95 GHz), primarily
undertaken with single-dish telescopes (e.g. Haschick et al. 1990;
Slysh et al. 1994; Val'tts et al. 2000; Ellingsen 2005) along with a
few smaller scale interferometric searches (e.g. Kurtz et al 2004;
Cyganowski et al. 2009). These have resulted in the detection of
about 200 class I maser sources in the Galaxy to date (Val'tts et
al. 2010).

Observations and theoretical calculations suggest that class I
methanol masers can form in the interface region between outflows
and the ambient molecular gas. The methanol abundance is
significantly increased in the shocked interface regions (e.g. Gibb
\& Davis 1998) and the gas is heated and compressed providing more
frequent collisions, which results in more efficient pumping
(Voronkov et al. 2006). Interferometric observations have shown the
location of some class~I methanol masers correlates closely with the
shocked gas in outflows as traced by 2.12 $\mu$m H$_{2}$ and SiO
(e.g. Plambeck \& Menten 1990 ; Voronkov et al. 2006). However, such
a close physical association between the shocks driven by outflows
and the class I masers has only been firmly established in a small
number of sources. One of the problems encountered in studying the
relationship between outflows (or shocks) and class I methanol
masers has been in finding an appropriate outflow tracer (one which
is frequently associated with class I maser emission). Recently
Cyganowski et al. (2008) have suggested that the 4.5~$\mu$m band of
the \emph{Spitzer Space Telescope's} Infrared Array Camera (IRAC)
offers a promising new approach for identifying massive young
stellar objects (MYSOs) with outflows. The strong, extended emission
in this band is usually thought to be produced by shock-excited
molecular H$_2$ and CO in protostellar outflows (e.g. Noriega-Crespo
et al. 2004; Reach et al. 2006; Smith et al. 2006; Davis et al.
2007; Ybarra \& Lada 2009, 2010). These extended 4.5~$\mu$m emission
features are commonly known as ``extended green objects'' (EGOs;
Cyganowski et al. 2008) or ``green fuzzies'' (Chambers et al. 2009)
due to the common color-coding of the 4.5~$\mu$m band as green in
three-color images. These objects are providing a new and powerful
outflow tracer for MSFRs. Cyganowski et al. (2008) have catalogued
over 300 EGOs from the Galactic Legacy Infrared Mid-Plane Survey
Extraordinaire (GLIMPSE) I survey (Churchwell et al. 2009), and they
divided cataloged EGOs into ``likely'' and ``possible'' MYSO outflow
candidates based primarily on the angular extent and morphology of
the 4.5~$\mu$m. Recent mid-infrared spectroscopic observations of
two EGOs identified by Cyganowski et al. showed the presence of
strong shocked H$_2$ in the 4-5 $\mu$m wavelength range from one of
the sources, but no evidence for shocked gas in the second (De
Buizer \& Vacca 2010). They suggest that some EGOs may be due to
spatial variations in the mid-infrared extinction and exaggerated
color stretches rather than shocked gas, however, the rate of
mis-classification of EGOs is at present very poorly determined.
Based on the mid-IR colors of EGOs and their correlation with
infrared dark clouds (IRDC) and known 6.7 GHz methanol masers,
Cyganowski et al. (2008) further suggests that most EGOs trace a
population of actively accreting MYSO outflow candidates. Utilizing
the published high resolution masers known at the time, Cyganowski
et al. found that 6.7 GHz class II methanol masers are associated
with 73\% (35 detections from 48 EGOs) of ``likely'' and 27\% (11
detections from 67 EGOs) of ``possible'' MYSO outflow candidate
EGOs.

Chen et al. (2009) have analyzed 61 EGOs from the Cyganowski et al.
catalog that have been included in class I methanol maser surveys.
Four previous class I methanol maser surveys (for the 44 GHz
$7_0$--$6_{1}$ A$^{+}$ transition by Slysh et al. (1994) and Kurtz
et al. (2004) and for the 95 GHz $8_0$--$7_{1}$ A$^{+}$ transition
by Val'tts et al. (2000) and Ellingsen (2005)) were included in
their statistical analysis. They found that 41 EGOs are associated
with one or both of the 95 and 44 GHz class I methanol masers within
1 arcmin, thus the expected detection rate of class I masers in EGOs
is 67\% at this resolution. Based on this high predicted detection
rate, they suggested that GLIMPSE-identified EGOs might provide one
of the best targeted samples for searching for class I methanol
masers. However their statistical study is also subject to unknown
influences produced by the target selection effects in the four
class I maser surveys they have utilised in the study. For example,
most of the currently known class I masers are associated with class
II masers (72\% as reported by Val\'tts \& Larinov 2007), for which
a good association with EGOs has already been demonstrated by
Cyganowski et al. (2008). Table 3 of Chen et al. (2009) shows that
most EGOs (57/61) used in their statistical analysis are also
associated with 6.7 GHz class II masers. According to Cyganowski et
al. (2008), the majority of EGOs (35/48=73\% in the likely sample)
are associated with 6.7 GHz class II masers, and the sample of known
class I masers is largely a subset of the sample of known class II
masers, so a high detection rate for class I masers towards these
EGOs is not unexpected. Recent 6.7 GHz class II and 44 GHz class I
methanol maser surveys toward $\sim$20 EGOs in the northern
hemisphere with the VLA reported by Cyganowski et al. (2009) also
show that $>$64\% and 89\% of EGOs are associated with class II and
class I methanol masers, respectively. However, their observations
only targeted a small number ($\sim$ 20) of ``likely'' outflow
candidate EGOs. And their 44 GHz survey sample is essentially a 6.7
GHz methanol maser selected sample (19 sources observed at 44 GHz
(17/19 with class I masers), of which 18 had associated 6.7 GHz
methanol masers (16/18 with class I masers)).

In order to determine whether there is indeed a high association
rate of class I methanol masers in EGOs and to investigate the
relationship between them, it is necessary to perform a class~I
maser search towards a full EGO-based target sample. In this paper,
we report a 95 GHz class I methanol maser survey towards an
EGO-selected sample which has been undertaken with the Australia
Telescope National Facility (ATNF) 22-m millimetre antenna at Mopra.
In Section 2 we describe the sample and observations, in Section 3
we present the results of the survey, the discussion is given in
Section 4, followed by our conclusions in Section 5.

\section{Source selection and Observations}
\subsection{Source selection}

The EGO catalog compiled by Cyganowski et al. (2008) lists a total
of 302 sources, of which 137 are classified as ``likely'' and 165 as
``possible'' outflow sources. All EGOs with declinations in the
range from -62$^\circ$ to +25$^\circ$ are accessible with Mopra.
However, there are some sources which have an angular separation of
less than $\sim$20$''$ (corresponding the half beam size of the
Mopra antenna; see below), from their nearest EGO. Where this
occurred, we observed at the position of those classified as
``likely'' outflow sources, or when neither fell into this category,
at the location with the larger 4.5 $\mu$m flux density. Chen et al.
(2009) reported that 61 EGOs have previously been observed in 44 and
95 GHz class I methanol maser surveys. For these sources we
re-observed 10 (of 16) EGOs which had not been observed at 95 GHz in
previous searches (see Table 1). Note that we did not make
observations of 6 of the sources without 95 GHz information in Chen
et al. sample in present survey (see Table 6 for this). After
applying these selection criteria we were left with a total of 191
EGOs to be observed. Among these sources, 65, 17 and 3 sources were
selected from Tables 1, 2 and 5 of Cyganowski et al. (2008),
respectively -- thus 85 ``likely'' outflow candidates; 55 and 51
sources were selected from Tables 3 and 4, respectively -- thus 106
``possible'' outflow candidates. In addition, a well studied MYSO
IRDC 18223-3 which shows distinctly extended 4.5 $\mu$m emission
(Beuther \& Steinacker 2007) was also included as a ``likely''
outflow candidate. In this paper we report the results of our 95 GHz
class I methanol maser survey toward these 192 sources.

Table 1 lists the target sample source parameters including the
source name (derived from the Galactic coordinates), the equatorial
coordinates and whether the EGO is associated with an IRDC, 6.7 GHz
class II methanol maser, 1665 or 1667 MHz OH maser, UC H{\sc ii}
region or 1.1 millimeter (mm) continuum source. The positional
accuracies of the 6.7 GHz methanol maser catalogs (Cyganowski et al.
2009; Caswell 2009; Xu et al. 2009; Caswell et al. 2010; Green et
al. 2010), OH maser catalog (Caswell 1998) and UC H{\sc ii} region
catalogs (Wood \& Churchwell 1989; Becker et al. 1994; Kurtz et al.
1994; Walsh et al. 1998; Cyganowski et al. 2009) used in our
analysis, are usually better than 1$''$. The positional uncertainty
of the 1.1 mm continuum sources in the BOLOCAM Galactic Plane Survey
(BGPS) is also of the order of several arcseconds (Rosolowsky et al.
2010). The {\em Spitzer} GLIMPSE point source catalog has also a
high positional accuracy (better than 1\arcsec), however, EGOs are
extended objects with angular extents between a few and $>$30$''$
(Cyganowski et al. 2008). Thus we have assumed an association
between EGOs and other tracers for separations of less than 30$''$.
For BGPS sources, we used their peak positions (rather than centroid
positions) available in the BGPS catalog (Rosolowsky et al. 2010)
and did not account for their sizes in the cross-matching. We list
the characteristics (including targeted/untargeted, area covered,
angular resolution, sensitivity and extent of overlap with the
GLIMPSE I survey area from which the EGO targets are drawn) for the
maser, UC H{\sc ii} region and BGPS 1.1 mm datasets used in our
analysis in Table 2. To more clearly show the associations among
these tracers and EGOs, we have also ploted the positions of the
masers, UC H{\sc ii} regions, and mm sources on the \emph{Spitzer}
3-color IRAC images for all 192 targeted sources in our Mopra survey
in Figure 1. From this figure, it can be seen that most of class II
methanol (marked by black crosses) and OH masers (marked by red
small circles) are close to the EGO positions (marked by blue
pluses) with typical separations between them of less than
5$\arcsec$. But for UC H{\sc ii} regions (marked by blue squares)
and mm sources (marked by yellow diamonds), they usually have a
slightly larger separation from the targeted EGO position. One
possible reason is that the UC H{\sc ii} regions and mm sources have
larger angular sizes, of order a few to a few tens of arcseconds.
Even though there are larger separations between EGOs and UC H{\sc
ii} regions and mm sources, we still consider them to be associated
in our analysis. However we recognize that in some cases there may
be no true physical association between them.

\subsection{Observations}

The observations were made using the Mopra 22m telescope during 2009
August 9-20. Mopra is located near Coonabarabran, New South Wales,
Australia. The telescope is at latitude 31 degrees south and has an
elevation of 850 metres above the sea level. A 3 mm Monolithic
Microwave Integrated Circuits (MMICs) receiver with a frequency
range from 76 to 117 GHz is installed. The UNSW Mopra spectrometer
(MOPS) provides four 2.2 GHz bands which overlap slightly to give
a total instantaneous bandwidth of 8 GHz. MOPS was used in zoom mode
for the observations reported here. In this mode, up to 4 zoom
windows can be selected within each 2.2 GHz band thus allowing up to
16 spectral lines to be observed simultaneously at high spectral
resolution. Each zoom window provides a bandwidth of 137 MHz with
4096 channels for each polarisation, which leads to a total velocity
range of $\sim$ 430 km s$^{-1}$ and a velocity resolution of 0.11 km
s$^{-1}$ per channel in the 3 mm band. The 8 GHz bandpass was
centred at 94.3 GHz to provide a complete coverage over the
90.3--98.3 GHz range. Within this range we configured the individual
zoom windows to simultaneously cover the strong lines in the 3 mm
band, e.g.  $1-0$ HNC (90.6635680 GHz), $5_{k}-4_{k}$ CH$_{3}$CN
($\sim$ 92 GHz), $1-0$ N$_{2}$H$^{+}$ ($\sim$ 93.17 GHz), $2-1$ CS
(97.9809533 GHz), $2-1$ C$^{34}$S (96.4129495 GHz), $8_{0}-7_{1}$
A$^{+}$ CH$_{3}$OH maser (95.1694630 GHz) and a series of lines of
$2-1$ CH$_{3}$OH ($\sim$ 96.7 GHz). In this paper we will focus on
the results of the 95 GHz class I methanol masers band. The results
from the other lines will be reported in subsequent publications.

Each source was observed in a position-switching mode with 1 minute
spent at the on-source position and 1 minute at a reference
position. This procedure was repeated a number of times to yield a
total of 11--13 minutes of on-source integration for most sources.
Two different reference positions offset by +8 and -8 arcminutes in
declination from the targeted EGO sites were used. The antenna
pointing was checked at hourly intervals by observing nearby bright
86 GHz SiO masers with known positions. The nominal pointing
accuracy is estimated to be better than 5\arcsec. The system
temperature was typically between 180-250 K depending on weather
conditions and telescope elevation, resulting in an rms noise level
of $\sim$30 mK per channel after averaging the two polarizations and
Hanning smoothing. The half-power main-beam size and the main-beam
efficiency were 36$''$ and 0.5, respectively at 3 mm band (Ladd et
al. 2005). The measured antenna temperatures (T$_{A}^{*}$) were
further calibrated on to a main beam temperature scale (T$_{MB}$) by
dividing by the main beam efficiency. Then a conversion factor of
9.3 Jy K$^{-1}$ can be estimated from Equ. 8.19 of Rohlfs \& Wilson
(2004) to convert the main beam temperature to a flux density. Thus
a direct conversion factor from T$_{A}^{*}$ to Jy would be 18.6 Jy
K$^{-1}$. The resulting flux density detection limit was $\sim1.6$
Jy (3 $\sigma_{rms}$; the typical rms noise is $\sim$0.55 Jy) per
channel. The sensitivity of 18.6 Jy K$^{-1}$ for our 95 GHz class I
methanol maser survey is significantly better than that of 40 Jy
K$^{-1}$ in previous similar surveys with Mopra (e.g. Val'tts et al.
2000; Ellingsen 2005), largely because the effective antenna
diameter has been increased from 15 to 22 meters in the 3~mm band
since 1999.

The spectral line data were reduced and analyzed with the ATNF
spectral line reducing software (ASAP) and the GILDAS/CLASS package.
During the reduction, quotient spectra were formed for each on/off
pair of observations which were then averaged together. A low-order
polynomial baseline fitting and subtraction, and Hanning smooth were
performed for the averaged spectra. For some sources the resulting
spectra still contained baseline ripples due to bad weather and for
these cases, we Fourier transformed the spectrum and flagged the
data in frequency space to reduce the influence of the baseline
ripple. This procedure was also undertaken with ASAP (for more
details see the Mopra
Website\footnote{http://www.narrabri.atnf.csiro.au/mopra/obsinfo.html}).
Usually the 95 GHz methanol spectra do not have a particularly
Gaussian profile, possibly because the spectra are blended with
multiple maser features within a similar velocity range. We
converted all the data into CLASS format and performed Gaussian
fitting of each possible maser feature for each detected source with
the CLASS package.

\section{Results}

\subsection{Nature of Detected Emission}

95 GHz class I methanol emission was detected towards 105 sites from
the 192 searched positions, yielding a detection rate of 55\% for
this survey. The detected sources are listed, along with the
parameters of Gaussian fits to their 95 GHz spectral features in
Table 3. The detected emissions range in strength from $\sim$ 0.2 to
108 Jy (corresponding to main beam temperatures T$_{BM}$ $\sim$ 0.02
to 11.6 K), derived from the Gaussian fits. Spectra of each of the
detected 95 GHz emission after Hanning smooth are shown in Figure 2.
The sources for which no 95 GHz emission was detected are listed in
Table 4, along with the 1 $\sigma$ noise of the observation
(typically less than 0.7 Jy). We consider a source to have been
detected if it exhibits emission stronger than 3 $\sigma$. All of
the 105 detected 95 GHz methanol sources are new detections in this
transition. Among them, only 13 sources have previously been
detected in the 44 GHz class I methanol transition in recent VLA
surveys (G34.26+0.15 detected by Kurtz et al. 2004; G10.29-0.13,
G10.34-0.14, G11.92-0.61, G18.67+0.03, G18.89-0.47, G19.01-0.03,
G19.36-0.03, G22.04+0.22, G23.96-0.11, G24.94+0.07, G25.27-0.43 and
G39.10+0.49 detected by Cyganowski et al. 2009). Therefore 92 new
class I methanol emission sources have been found in this survey.
From the spectra of the class I methanol emission sources shown in
Figure 2 it can be seen that the class I emission often consists of
one or more narrow spectral features which can be treated as maser
emission, apparently superimposed on broader emission features.
Examination of Table 3 shows that spectral features with widths $>$
1 km s$^{-1}$ are common in the detected sources. The characteristic
of broader emission features in class I transition is very similar
to that reported by Ellingsen (2005). From single-dish observations
we cannot determine whether these broad features are
quasi-maser/quasi-thermal emission, or due to blending of a number
of weaker narrow maser features (since the Mopra spectra almost
certainly include multiple class I maser spots blended within the
beam, multiple weak masers at different velocities could in
aggregate produce some of the broad emission features seen). However
broad, quasi-thermal like spectral profiles are commonly observed in
class I methanol masers (see for example Kurtz et al. 2004, or
Voronkov et al. 2006). The majority of the broad emission in these
sources when observed interferometrically is observed to be maser
emission. Thus, although we cannot conclusively rule out the
possibility that some of the sources may have thermal emission
contributions, comparison with previous observations makes us very
confident that the majority of the broad emission is maser emission.
Moreover some of the EGOs with previous 44 GHz maser detections
(Kurtz et al. 2004; Cyganowski et al. 2009) also show similar broad
emission profiles at 44 GHz (for Cyganowski et al. 2009, the 44 GHz
maser spectra were not directly presented, but they can be
determined from the fitted intensities of the 44 GHz emission listed
in Table 8 of their work). The similar broad emission features from
class I methanol masers associated with EGOs are also detected and
spatially unresolved in 44 GHz VLA observations (Kurtz et al. 2004;
Cyganowski et al. 2009), which suggests that they are generally
masers. It has yet to be confirmed through interferometric
observations that the broad 95 GHz spectral features are also maser
emission, although given the results of numerous past
interferometric observations of class I masers this appears likely
for the majority.

We also checked the flux densities of 44 GHz maser emission for the
13 EGOs which were observed as 95 GHz masers in our Mopra survey,
and found that the peak flux densities of the 44 GHz masers are
typically 2-3 times stronger than that of 95 GHz masers, consistent
with the statistical result from Val'tts et al. (2000). However the
primary beam of the VLA (1$\arcmin$ at 44 GHz) is larger than that
of Mopra (36$\arcsec$ at 95 GHz). Thus the detection of stronger 44
GHz emission may be because for some sources with extended maser
emission the VLA can detect additional maser emission which is
within the VLA primary beam, but outside the Mopra beam (see
Cyganowski et al. 2009). To account for this potential affect, we
compared only the 8 sources with all 44~GHz maser emission detected
by the VLA locating in a compact region and within Mopra beam. We
found that the peak flux densities of 44 GHz emission are still 2-3
times stronger than that of 95 GHz emission for these sources.

In Table 3 we also list the distance and the integrated maser
luminosity of each of the 105 detected methanol maser sources. The
distance was estimated from the Galactic rotation curve of Reid et
al. (2009), with the Galactic constants, R$_{\odot}$= 8.4 kpc and
$\Theta_{\odot}$= 254 km s$^{-1}$. Class I masers are generally
found near the V$_{LSR}$ as measured from the thermal gas (e.g.
Cyganowski et al. 2009). We adopted the velocity of the brightest
feature in the 95 GHz maser spectrum (listed in column 3 of Table 2)
in the distance calculation. The near kinematic distance was adopted
for sources with a near/far distance ambiguity. Given that EGOs are
by definition extended sources, there will be bias towards nearby
sources in their identification and also the nearby sources will be
brighter and easier to detect. Another argument for EGOs being more
likely at the near kinematic distance is their associations with
IRDCs, because the identification of IRDCs is greatly biased toward
the near kinematic distance (see Jackson et al. 2008). Thus the near
distances adopted for these sources are likely to be more reasonable
for most sources. The distances to G49.07-0.33, G309.91+0.32,
G310.15+0.76 which could not be derived from the Galactic rotation
curve were assumed to be 5 kpc.

\subsection{Detection Rates}

The results of our single-dish survey prove that there is indeed a
high detection rate ($\sim$55\%) of class I methanol masers towards
EGOs. It expands the number of published class I methanol masers by
about 50\% (an additional 92 on top of the 198 listed by Val'tts et
al. 2010). However given the large beam size of Mopra and clustered
massive star formation, a detection in a pointing towards an EGO
does not necessarily mean that the detected class I maser is truly
associated with the EGO, though recent VLA observations towards
$\sim$20 EGOs reported by Cyganowski et al. (2009) revealed that
nearly all 44 GHz class I masers trace the diffuse green 4.5 $\mu$m
features of EGOs except for one case (G28.28-0.36). If this is true
of the sample as a whole then it suggests that the false association
rate between EGOs and detected class I masers is likely to be very
low with a range of a few to (at most) 10\%. Where there were
multiple EGOs included in the large beam of Mopra, we pointed at the
position of the EGO classified as a ``likely'' outflow source, or
where there were two such sources, at the position with the largest
4.5 $\mu$m flux density (see Section 2.1). It is not clear which of
the EGOs (all, or only one) included in such observations are
associated with the detected maser emission. Thus it requires
further high resolution observations toward the detected class I
masers to determine whether they are truly physical associations
with EGOs or which EGOs are associated with them. In subsequent
analysis, we have assumed that the detected maser emission is
associated with the EGOs used as the targeted point.

The detection rate of 55\% achieved in this survey is slightly lower
than the predicted value of 67\% from our previous statistical
analysis (Chen et al. 2009). One possible reason for this is that
our previous analysis combined 44 and 95 GHz class I maser searches,
and emission from the 44 GHz transition is generally 3 times
stronger than that at 95 GHz (Val'tts et al. 2000). Thus it may be
possible to detect a significant number of additional sources if we
were to make a search with similar or better sensitivity at 44 GHz.
To test this conjecture, we checked all 20 EGOs with detected 44 GHz
masers in previous surveys (Kurtz 2004 and Cyganowski et al. 2009),
and found that only 13 EGOs have detected 95 GHz emission at our
Mopra survey sensitivity of $\sim$2 Jy (i.e. the detection rate of
95 GHz methanol masers in these EGOs is 13/20$\sim$65\%). All the
remaining 7 EGOs detected at 44 GHz for which we did not detect 95
GHz emission have peak flux densities of less than 6 Jy for the 44
GHz masers. This is consistent with our expectation that these 7
EGOs might not be detected class I masers with the current detection
limit (corresponding to $\sim$ 6 Jy at 44 GHz when considering the
44/95 GHz correlation from Val'tts et al. 2000). If we extrapolate
this result to our survey it suggests that we have detected around
2/3 of the true number of class I maser sources in our search. If
this were correct then our detection rate would be well in excess of
the prediction of Chen et al. (2009). This extrapolation naively
ignores the likely presence of selection biases in the VLA
comparison sample and the effect of the different primary beam of
the VLA and Mopra as stated above, but these are not readily
assessed. The true rate of association of class I methanol masers
with EGOs can perhaps only be accurately determined through targeted
sensitive 44 GHz observations towards the non-detected sources from
this survey.

\subsection{Associations with Class II Methanol Masers}

Table 1 gives information on the association between EGOs and class
II methanol masers. For those sources in the Galactic longitude
range covered by sections of the MMB survey published to date
(Caswell et al. 2010; Green et al. 2010), or other similarly
sensitive observations (Caswell 2009, Cyganowski et al. 2009, Xu et
al. 2009) we have used that data. For the remaining EGOs we have
made additional 6.7 GHz observations with the Mt Pleasant 26m
telescope. The Mt Pleasant observations will be described in detail
in a separate publication, but the basic characteristics are
summarised here. At a frequency of 6.7 GHz the Mt Pleasant 26m
telescope has a 7$\arcmin$ beam FHWM and a system equivalent flux
density (SEFD) of around 800~Jy in each of two orthogonal circular
polarizations. The correlator was configured with 4 MHz bandwidth
and 4096 spectral channels for each circular polarization yielding a
velocity range for the observations of approximately 180 km s$^{-1}$
and a velocity resolution (after Hanning smoothing) of 0.09 km
s$^{-1}$. Observations were made using 10 minute integrations
towards each EGO target position observed at 95 GHz and the 3
$\sigma$ detection limit of the observations is around 1.5 Jy,
comparable to that achieved for the 95 GHz Mopra search. Although it
is possible that some of the EGOs not detected in the Mt Pleasant
observations may have a weak associated 6.7 GHz maser, this is
likely to effect only a small number of sources. From comparison
with the MMB observations we estimate this to be less than 10\% of
non-detections. While the class II methanol maser data used in our
comparison with the class I masers are not derived from a
homogeneous set of observations, their relative sensitivities differ
only by approximately a factor of 2 and are unlikely to introduce
significant additional uncertainty into our statistical analysis.

We have compared the cataloged EGOs with the subset of class II
methanol masers for which accurate positions have been published
(see Section 2.1). Of the 192 cataloged EGOs, 49 are found to be
associated with class II masers and 81 were not. The results for the
EGOs which are not associated with class II masers were primarily
determined from the surveys for 6.7 GHz methanol masers with the Mt
Pleasant 26m (see above). The remaining 62 EGOs were detected 6.7
GHz methanol masers by the Mt Pleasant 26m telescope, but lack a
high-precision position, thus we consider them as sources for which
we have ``no'' information with respect to the class II masers in
our work. Among the 105 EGOs with detected class I methanol masers,
we found that 39 EGOs are, and 31 EGOs are not associated with class
II methanol masers, respectively. For the other 35 EGOs we have
insufficient information on the class II masers due to the lack
accurate position observations of the class II masers for these
sources. The sources associated with only class I methanol masers,
but without class II methanol masers are especially important for
our understanding of the properties of class I methanol masers,
because most previous class I methanol maser surveys were targeted
towards known class II methanol masers. We have undertaken further
analysis and discussion of this issue in Section 4.2.

On the basis of one of the earliest class I methanol maser searches
to detect a large number of sources, Slysh et al. (1994) suggested
that there exists an anti-correlation between the flux densities of
class I and class II methanol masers towards the same sources.
Ellingsen (2005) further investigated this finding for a sample of
class I methanol masers detected towards a statistically complete
sample of class II masers and in that case found no evidence for an
anticorrelation. We have also compared the peak flux densities of
class I and class II methanol masers in 39 sources associated with
both class masers. The peak flux densities of the class II masers
are from interferometric studies (Cyganowski et al. 2009; Caswell
2009; Xu et al. 2009; Caswell et al. 2010; Green et al. 2010). The
logarithms of the peak flux densities of class I masers vs. that of
class II masers are shown in Figure 3. From this figure it can be
seen that there is no statistically significant correlation (or
anticorrelation) between the flux densities of class I and class II
methanol masers in our observed EGO sample, consistent with the
result of Ellingsen (2005). The flux densities of the class II
methanol masers used in our comparison were all obtained from
interferometric studies, whereas the previous comparisons of Slysh
et al. (1994) and Ellingsen (2005) used single-dish data. However,
the class II methanol maser emission is usually distributed over a
compact region typical $<$0.2$''$ and are not resolved by connected
element interferometry (e.g. the VLA). This means that their flux
density is the same in interferometric and single dish studies and
hence this difference produces no new or additional biases to our
statistical comparison.

\subsection{Associations of Class I methanol masers with other star formation tracers}

In this section, we compare the detection rates of class I methanol
masers in different subsamples, including ``likely'' and
``possible'' outflow candidates, IRDC and non-IRDC, those associated
with class II methanol masers, OH masers, and UC H{\sc ii} region
subsamples, etc. In order to more clearly compare the various
samples we have listed the information relating to the class I
methanol maser detections for all 192 sample sources in the last
column of Table 1. The detection rates for each of the different
categories are summarized in Table 5. It should be remembered that
the detection rate in some categories will be affected by the
limitations of single dish surveys with a large beam size and the
clustering which occurs in high-mass star formation regions (as
described in Section 3.1).

A total of fifty three 95 GHz methanol masers were detected towards
the 86 likely outflow sources targeted by our observations. The 53
detections include 40, 10 and 2 sources from Tables 1, 2 and 5 of
Cyganowski et al. (2008), respectively, and also IRDC 18223-3. This
corresponds to a detection rate of 62\% for the ``likely'' outflow
sources. The remaining 52 class I masers detections were made
towards the 106 ``possible'' outflow candidate EGOs observed. For
the possible outflow sources, 27 and 25 were from Tables 3 and 4 of
Cyganowski et al. (2008), respectively. This corresponds to a
detection rate of 49\%. A {\em z}-test finds that the difference in
these detection rates is significant at the 90\% level (i.e. it is
marginally significant). Regardless of this, it is apparent that the
detection rate of class I methanol masers in the ``likely'' outflow
subsample is only slightly higher than that in the ``possible''
outflow subsample. This suggests that the class I methanol maser
emission may be not very sensitive to the outflow classifications
(i.e. ``likely'' and ``possible'') seen from the IRAC images.
Alternatively, if we assume that our finding for ``likely'' outflow
sources is the true rate of association between EGOs and class I
methanol masers then this suggests that approximately 49/62=79\% of
the ``possible'' outflow candidates are indeed outflow sources.

Dividing our sample of EGOs searched for class I methanol masers on
the basis of their association (or otherwise) with IRDCs, we found
that 71 of the 128 EGOs associated with an IRDC exhibit class I
methanol maser emission (a detection rate of 55\%). Whereas, 34 of
the 64 EGOs without an IRDC were detected as class I methanol maser
sources (a detection rate of 53\%). IRDCs are generally thought to
host an early stage of the high-mass star formation process, so it
is somewhat surprising that our results show no difference between
the detection rate of class I methanol maser in those EGOs which are
and are not associated with an IRDC. The visibility of an IRDC is
dependent on both the strength of the mid-infrared background
emission and the amount of foreground emission, particularly at 8
$\mu$m (Cyganowski et al. 2008; Peretto \& Fuller 2009). If there is
no, or weak 8 $\mu$m background emission in a particular region, an
IRDC may not be apparent even where dense molecular gas and very
young MYSOs are present. Moreover, MYSOs and YSOs of a range of
masses and evolutionary states are also found in IRDCs (e.g. Pillai
et al. 2006; Ragan et al. 2009; Rathborne et al. 2010). So sources
not associated with IRDCs do not necessarily host a later
evolutionary stage than those which are.

Our target EGO sample included 49 sources associated with known
class II methanol masers for which the position is accurately known
and 81 sources which have been searched for class II methanol maser
emission with no detection by the Mt Pleasant 26 m telescope (see
Table 1 and Section 3.2). We found that 39 of the 49 sources
associated with class II methanol masers were detected in the 95 GHz
class I methanol transition. Thus there is a very high detection
rate ($\sim$ 80\%) of class I methanol masers towards EGOs which are
also associated with class II methanol masers, which is somewhat
lower than the 89\% reported by Cyganowski et al 2009, but likely
not significant given the sensitivity differences between the
observations. In comparison only 31 of the 81 sources not associated
with a class II methanol maser were found to have an associated 95
GHz class I methanol maser (a detection rate of 38\%). A more
in-depth discussion about the lower detection rate of class I
methanol masers towards those EGOs without an associated class II
methanol maser and the evolutionary relationship between class I and
class II methanol masers is given in Section 4.2.

Comparing the positions of the EGOs we observed with the spatial
region of the Parkes/ATCA 1665/1667 MHz OH maser survey (see Table 2)
(Caswell 1998), we found that there were 104
targeted EGO sources within the survey area of the Parkes OH maser
survey. Amongst these, 14 EGOs were, and 90 EGOs were not associated
an OH masers. Nearly all (13 of 14) of the EGOs associated with an
OH maser, while approximately one-half (43 of 90) of the sources not
associated with an OH maser were found to have a class I methanol
maser.

We also note that there is a very high detection rate (11/13=85\%)
of class I methanol masers towards those EGOs which are associated
with an UC H{\sc ii} region. For most EGOs no deep centimeter
continuum data is available. There are two H{\sc ii} region survey
datasets that we have used to compile a category of ``without
associated UC H{\sc ii} region'' for comparison. The survey of
Becker et al. (1994) covered the region of $|b|<0.4^\circ$ and
$l=350^\circ-40^\circ$ (see Table 2). The other observations are
those by Cyganowski et al. (2009) towards $\sim$20 EGOs in the
northern hemisphere, who found that no 44 GHz continuum emission was
detected toward 95\% of their surveyed EGO sample. Using the
published data from these two sets of observations, we compiled a
sample of 34 EGOs that are not associated with an UC H{\sc ii}
region. From this sample, 21 sources were found to have an
associated class I maser. Thus the detection rate of class I masers
of 21/34=62\% towards EGOs without associated UC H{\sc ii} regions
is lower than that in sources which are associated with UC H{\sc ii}
regions. The size of the subsample associated with UC H{\sc ii}
regions used in our statistical analysis is small, and they may be
biased since many of the large-scale UC H{\sc ii} region surveys
cited were targeted based on IRAS colors. In addition, high-mass
star formation usually occurs in a cluster environment, so for an
EGO associated with an UC HII region, it is not clear a priori
whether the 4.5 $\mu$m outflow is driven by the UC H{\sc ii} region
or by another (potentially lower-mass or less evolved) source in a
(proto)cluster. However, given that UC {\sc ii} regions are
relatively rare towards EGOs, the high detection rate of class I
methanol masers in those few sources where they have been observed
to date is suggestive.

There are 63 EGOs which fall within the 1.1 mm BGPS survey area
(Rosolowsky et al. 2010). Fifty four of these are associated with
BGPS sources within 30$''$, and 9 are not (see Table 2). We find
that the detection rate of class I maser in the EGOs with an
associated BGPS source (35/54=65\%) is higher than that in those
without an associated BGPS source (1/9=11\%; only one EGO
G34.39+0.22 in this category was detected to have class I maser
emission). Conversely, all class I maser sources except G34.39+0.22
which fall within the BGPS survey region have an associated 1.1 mm
BGPS source.

\section{Discussion}
\subsection{The mid-IR colors of the class I methanol masers}

Based on their mid-IR colors ([3.6]-[5.8] versus [8.0]-[24]),
Cyganowski et al. (2008) have suggested that most EGOs fall in the
region of color-color space occupied by the youngest MYSOs and are
surrounded by substantial accreting envelopes (see Figure 13 in
their work). We have performed additional color-color analysis for
EGOs to further investigate the distinguishing mid-IR properties of
EGOs with and without an associated class I methanol masers. Our
mid-IR color analysis includes the 192 EGOs targeted for our class I
maser observations and an additional 51 EGOs that are listed in our
previous work (Chen et al. 2009), but were not observed in our Mopra
survey. We list the information (including their associations with
class I masers and high-precision position class II masers) of these
51 EGOs in Table 6. There are 6 EGOs without 95 GHz class I maser
information listed in this table since they were omitted in our
Mopra survey (see also Section 2.1). Adopting the integrated mid-IR
flux densities in the four IRAC bands presented in Tables 1 and 3 of
Cyganowski et al. (2008), we have plotted a diagram of the
[3.6]-[4.5] versus [5.8]-[8.0] colors of these selected EGOs in
Figure 4 (note that we do not consider the flux density limitation
on each IRAC band denoted by column 12 of Tables 1 and 3 of
Cyganowski et al. in our analysis). In total, 81 and 58 EGOs with
and without an associated class I methanol maser (see Tables 1 and
6) are shown in this figure represented by red and blue triangles,
respectively. The regions of color-color space for sources at
different evolutionary Stages I, II and III, derived from the 2D
radiative transfer model by Robitaille et al. (2006) are also marked
in Figure 4. We found that most EGOs fall in the region occupied by
the youngest protostar models (Stage I), consistent with the
conclusions from Cyganowski et al. (2008). There is significant
overlap in colors between sources with and without class I methanol
masers. Figure 4 also shows that many EGOs lie in the upper-left of
the color-color diagram, and outside the Stage I evolutionary zone.
One possible reason for this is that the colors of these sources are
effected by reddening. The reddening vector for A$_v$$=$20 derived
from the Indebetouw et al. (2005) extinction law is shown in Figure
4. An A$_v\sim$80 can produce reddening of approximately 1.4 mag in
the [3.6]-[4.5] color, which would be sufficient to return most of
these source to the Stage I region. However, such a large extinction
value A$_v$$\sim$80 is of dubious plausibility, since the path to
the Galactic center has a total A$_v$$\sim$25 (Indebetouw et al.
2005). Another possibility is that because the mid-IR flux density
measurements were determined from an extended region (i.e. extended
green region), with a typical scale of a few to 30$''$, they may
include emission from many GLIMPSE point sources which are not
physically associated with the EGO, shifting the colors for some
sources outside the Stage I region. Moreover the integrated fluxes
likely include emission mechanisms (H$_{2}$ and PAH line emission in
particular) which were not included in the Robitaille et al. models.
This also may result in some sources with extended 4.5 $\mu$m and
PAH emission lying outside the Stage I region.

We have also undertaken additional similar color-color analysis
using flux measurements for all EGOs extract from the highly
reliable GLIMPSE point source catalog (rather than the less reliable
GLIMPSE point source archive). This analysis may allow us to
determine whether it is possible to refine the criteria for
targeting class I methanol masers using mid-IR colors of GLIMPSE
point sources, similar to the analysis undertaken by Ellingsen
(2006) for class II methanol masers. The sample includes the 192
EGOs in the current Mopra observations and an additional 51 EGOs
listed in our previous work (Chen et al. 2009). Although EGOs are by
definition extended objects, the GLIMPSE point catalog allows us to
study the characteristics of the possible driving source of the EGOs
and class I methanol masers. To decrease the contamination of our
investigation of the EGO driving sources from chance associations,
we assumed that the driving source is the closest point source to
the cataloged EGO position (within 5$\arcsec$), with flux
measurements in all four IRAC bands. Using these criteria we
identified the assumed driving GLIMPSE point sources for 126 EGOs
(including 74 associated with detected class I methanol masers and
52 without class I methanol masers; see Tables 1 and 6). In Figure 5
we have marked these closest associated GLIMPSE point sources using
red and blue triangles for EGOs which are and are not with
associated class I methanol masers respectively. Examination of
Figure 5 clearly shows that the GLIMPSE point sources which lie
closest to the EGOs are predominantly inside the color-color region
representative of Robitaille et al. evolutionary Stage I.
Qualitatively Figures 4 and 5 show similar color-color
distributions, except perhaps for a greater spread in the
[3.6]-[4.5] color for the point source data compared to the
integrated fluxes. We have used the sample of 126 GLIMPSE point
sources identified in the manner outlined above to perform our
subsequent color-color analysis for EGOs with/without class I
methanol masers. It is almost certain that there are some false
associations between the GLIMPSE point sources assumed to be the
driving sources and the true driving sources of the EGOs. For
example, in some cases the true driving sources for EGOs may not
appear in the GLIMPSE point source catalog, either due to
saturation, the presence of bright diffuse emission (which limits
point source extraction), or inherently extended morphology in the
IRAC bands with the IRAC resolution, e.g. due to extended PAH
emission or extended H$_{2}$ emission from outflows (see Robitaille
et al. 2008, Povich et al. 2009, and Povich \& Whitney 2010).
Moreover the EGO position cataloged by Cyganowski et al. (2008)
adopted the position of the brightest 4.5 $\mu$m emission within the
extended region of the EGO, but it is not clear that the brightest
4.5 $\mu$m emission associated with an EGO must necessarily lie
close to the driving source. However, given the extended nature of
EGOs such identifications will always be somewhat problematic. The
presence of false associations will add confusion to attempts to
identify any color-color differences between EGOs with and without
class I methanol masers; however, provided the mis-identification
rate is not too large (it is not possible to make a quantitative
estimate for this at present) it is unlikely to mask the difference
completely, if it is present.

We find that the color-color regions occupied by GLIMPSE point
sources which are and are not associated with class I methanol
masers are not significantly different. For both groups they
predominantly lie in a box region -0.6$<$[5.8]-[8.0]$<$1.4 and
0.5$<$[3.6]-[4.5]$<$4.0. This color-color region is very similar to
that occupied by 6.7 GHz class II methanol masers identified by
Ellingsen (2006). But interestingly, we note that the class I
methanol masers extend to smaller [3.6]-[4.5] colors than do the
class II methanol masers for which the [3.6]-[4.5] color is usually
greater than 1.3.

Similar to what is seen in Figure 4, Figure 5 shows that many
GLIMPSE point sources which are closest to an EGO lie in the
upper-left of the color-color diagram, and outside the Stage I
evolutionary zone derived from 2D radiative transfer models
(Robitaille et al. 2006). Some of the GLIMPSE point sources have
redder [3.6]-[4.5] colors ($>$3), compared to those seen in Figure 4
which typically have [3.6]-[4.5] $<$ 3. As discussed above, these
sources may be those which suffer larger reddening, however it
requires extreme extinction (A$_v >100$) to produce such large color
shifts ($>$2) in [3.6]-[4.5]. The typical A$_v$ of an IRDC simply
estimated from the parameters provide by Peretto \& Fuller (2009) is
25 (here we adopted an average optical depth at 8 $\mu$m of 1.15 for
IRDCs, and A$_{8\mu m}$/A$_v$=0.045 from that work). Since most EGOs
are associated with IRDCs (Cyganowski et al. 2008), this shows that
typical A$_v$ for IRDCs can not account for the entirety of redder
colors of these sources.   From our calculations of A$_{v}$ for the
1.1 mm BGPS sources associated with EGOs which have class I maser
detections (see Table 7, with more details described in Section
4.3), the estimated A$_{v}$ of the 1.1 mm clumps associated with
EGOs outside stage I region ranges from 10 to 40 (with an average of
20), similar to that for sources inside the Stage I region which
range from 5 to 50 (with an average of 23). This also suggests that
the reddening does not play an important role in explaining the
redder colors of these sources. Another possible reason for this is
that the GLIMPSE point source photometry (similar to the integrated
flux measurements discussed above) may be affected by contributions
from extended H$_{2}$ emission, while the classifications of Stage
I-III by Robitaille et al. (2006) did not consider these emission
mechanisms. The EGOs associated with GLIMPSE point sources outside
Stage I region have a significantly higher detection rate for class
I methanol masers (21/28=75\%; see Tables 1 and 6) than that
observed in the full sample (55\%). We have also checked the masses
of 1.1 mm BGPS sources associated with EGOs which have class I
methanol maser detections (see Table 7, and the additional details
described in Section 4.3), and found that the mass range of the BGPS
sources associated EGOs outside stage I region (covering the range
1000-6000 M$_{\odot}$ with an average of 2500 M$_{\odot}$) is
significant higher than those associated with EGOs in Stage I region
(73-2000 M$_{\odot}$, with an average of 1200 M$_{\odot}$). Thus
they may correspond to MYSOs with an extremely high mass envelope
which is more deeply embedded causing redder colors. Moreover, the
detection rate of class I methanol maser in EGOs tends to be higher
(30/40=75\%) in the color-color region with [3.6]-[4.5] $>$ 2.4.
Given that EGOs are identified and defined by their excess 4.5
$\mu$m emission, the high detection rate of class I masers towards
EGOs that fall in the left-upper regions make sense if the excess
4.5 $\mu$m emission is due to shocked H$_{2}$ in outflows, and so
sources in the left-upper region may have particularly strong/active
outflows which can readily produce maser emission. This suggests
that GLIMPSE point sources with redder [3.6]-[4.5] color are the
best target population for class I methanol maser searches. However,
the small number ($\sim$30) of class I methanol maser and small
number ($\sim$40) of GLIMPSE point sources in this region of color
space should be taken into account when drawing any conclusions.

\subsection{An evolutionary sequence for class I and II masers}

Ellingsen et al. (2007) suggested that the common maser species
(class I and II methanol, water, and OH masers) may help identify
the evolutionary phase of a high-mass star, and proposed a possible
evolutionary sequence for these common maser species. This proposed
sequence has recently been refined and quantified by Breen et al.
(2010a) in their Figure 6. However, there remains significant
uncertainty about where within star formation regions the different
maser species arise and the evolutionary phase they are associated
with. In this work, we focus on the evolutionary sequence for class
I and II methanol masers. In previous work, one of the main
difficulties in determining the relative evolutionary sequence for
class I and II masers has been the lack of a large sample of sources
associated with class I masers but not class II masers. Ellingsen
(2006) investigated the mid-IR colors of the associated GLIMPSE
point sources for a relatively small sample of class II methanol
masers associated with and without class I methanol masers numbering
$\sim$10 for each group, and found there is a tendency for the
sources with an associated class I methanol maser to have redder
GLIMPSE colors than those without class I methanol masers. Based on
the assumption that the redder colors are associated with more
deeply embedded and hence youngest stellar objects, Ellingsen (2006)
suggested that some class I methanol masers may precede the earliest
class II methanol maser evolutionary stage. However, the absence of
a comparison sample of class I methanol masers with no associated
class II methanol masers presents a significant limitation to the
Ellingsen (2006) work. Our class I methanol maser survey towards
EGOs has identified 31 sources which are associated with class I,
but not class II, methanol masers.

To test the proposed evolutionary scenario for class I and II
methanol masers, the EGOs were split into three subsamples on the
basis of which class methanol masers they were associated with (see
Tables 1 and 6): 1) associated only with class I methanol masers (32
members in total; 31 from our surveyed sample and 1 from Chen et al.
2009 sample); 2) associated only with class II methanol masers (20
members in total; 10 from our surveyed sample and 10 from Chen et
al. 2009); 3) associated with both class I and class II methanol
masers (72 members in total; 39 from our surveyed sample and 32 from
Chen et al. 2009). In compiling the second and third subsamples we
only considered sources for which the position of the class II maser
emission is known to high accuracy (i.e. the sources with class II
methanol maser information marked by ``Y'' in Tables 1 and 6). IRAC
and Multiband Imaging Photometer for Spitzer (MIPS) 24 $\mu$m colors
provide a diagnosis for YSO evolutionary state (Robitaille et al.
2006). We plot [3.6]-[5.8] versus [8.0]-[24] color diagram using the
flux measurements from Tables 1 and 3 of Cyganowski et al. (2008)
for the above three subsamples (note that we do not consider the
flux density limitations on the IRAC and MIPS bands in this plot) in
Figure 6 with different symbols. The regions of color-color space
for sources at different evolutionary stages I, II and III derived
from Robitaille et al. (2006) are also marked in Figure 6. In total,
we have 26 EGOs containing both class I and II methanol masers, 7
EGOs associated with only class II methanol masers and 25 EGOs
associated with only class I methanol masers in this figure.
Comparing the color distributions of these three subsamples with the
color-color space occupied by the evolutionary stages derived from
Robitaille et al. (2006), we find that all class II maser EGOs
(including both class I and II maser subsample and the only class II
maser subsample) are located in the region of Stage I, i.e. the
easiest evolutionary stage, while all but one class I maser only
EGOs are also located in the Stage I region (the exception is
G317.88-0.25 which lies in Stage II). However, as seen in Figure 6,
despite the significant overlap of the various subsamples, EGOs
which are associated with only class I methanol masers extend to
less red colors than those associated with only class II methanol
masers and both class I and II methanol masers.

Here we propose a number of possible explanations as to why the
mid-IR sources associated with only class I methanol masers have
less red colors than those associated with class II methanol masers:

1. The EGOs associated with only class I maser are less heavily
extincted than those associated with class II masers.

2. The stellar mass range of objects with associated class I
methanol masers extends to lower masses than that of objects with
class II methanol masers. The lower mass sources may be generally
less deeply embedded and hence have a less red colors, than the
higher mass objects.

3. There may be two epochs of class I methanol maser emission
associated with high-mass star formation. An early epoch which
overlaps significantly with the class II methanol maser phase and a
second phase which occurs after class II methanol maser emission has
ceased.

The first possibility is supported by the evidence that the mid-IR
color differences among the three subsamples shown in Figure 6 are
mostly along the direction of the reddening vector. But it needs a
very large A$_v$ to produce the color shifts observed between the
EGOs associated with only class I masers and those associated with
class II masers (e.g. A$_v\sim$80 corresponds to a color shift of 3
in [8.0]-[24]). Even though at present we can not accurately
determine A$_v$ for our full sample sources, as discussed in Secion
4.1 from estimations of A$_v$ for a few sources associated with 1.1
mm BGPS listed in Table 7 and the typical A$_v$ for IRDCs which are
often associated with EGOs, it appears that reddening alone is
unlikely to be responsible for the observed color differences among
the three subsamples.

The second possibility is also consistent with other recent
observations of class I methanol masers. Class I methanol masers are
known to be associated with some regions that are forming only
low-mass stars (Kalenskii et al. 2006, 2010). This suggests that
class I methanol masers can be associated with lower stellar mass
sources than class II methanol masers and hence supports this
possibility. To further test this we have compared the detected 95
GHz class I maser luminosity distributions of four subsamples. The
subsamples of class I masers are (a) those not associated with class
II masers; (b) those associated with class II masers for which an
accurate position has been measured; (c) those associated with an UC
H{\sc ii} and (d) those associated with an OH maser. The
distribution of the 95 GHz class I maser luminosity for each of
these subsamples is shown in Figure 7. This shows that most
(25/31$\approx$80\%) of the sources which are not associated with
class II methanol masers are located in the lowest luminosity bin
(less than 5$\times$10$^{-6}$ L$_\odot$); whereas those associated
with class II masers, UC H{\sc ii} regions and OH masers have a
relatively small fraction (typically 40\%) in this lowest luminosity
bin. One explanation for the observed distributions is that class I
masers can be associated with lower stellar mass sources than class
II masers or the other two tracers (OH and UC H{\sc ii}), since
class I maser excited by outflows from low stars are expected to be
less luminous. Recalling Section 3.3, the detection rate of class I
methanol masers in the sources without an associated class II maser
(37\%) is lower than that in the sources with an associated class II
maser (80\%). This statistical result also seems to support the
hypothesis that class I methanol masers can extend lower stellar
mass sources since less luminous class I masers excited from lower
mass stars are harder to detect with the same sensitivity. This
hypothesis is also supported by our further analysis of the
relationship between class I methanol masers and 1.1 mm BGPS sources
(discussed in Section 4.3).

The last hypothesis is more speculative, as it requires the
mechanism through which class I methanol masers are produced to be
switched off and then at a later time on again. It is generally
considered that sources with associated OH masers and UCH{\sc ii}
regions lie towards the later evolutionary phases (Breen et al.
2010a). The very high rate of association we have found for class I
methanol masers towards OH masers (93\%; see Table 5) demonstrates
that some class I methanol masers may be present at these later
stages. However the current single dish survey with large beam size
is not sufficient to argue that the driving source of OH maser is
also responsible for exciting the class I maser emission. Further
high-resolution observations are needed to definitively establish
whether the class I masers are truly associated with the same MYSO
as the OH masers, although the results of Cyganowski et al. (2009)
suggests that most of the detected class I masers will be associated
with the targeted EGOs. Recently Voronkov et al. (2010) presented
new high resolution observations which strengthen the case that some
class I methanol masers are produced in shocks driven into molecular
clouds from expanding H{\sc ii} regions. The 9.9 GHz class I
methanol masers (detected towards 4 of 48 class I maser sources
observed by Voronkov et al.), are all associated with relatively old
sources, e.g. H{\sc ii} regions and OH masers. They also tentatively
report a detection rate of greater than 50\% for 44 GHz class I
methanol masers towards OH masers which are not associated with
class II methanol masers. This indicates that the class I masers can
extend beyond the time when class II masers are destroyed, and
overlap well into the time when OH masers are active. Voronkov et
al. suggest that these findings are consistent with the cloud-cloud
collision hypothesis for class I methanol masers which has been
realized in some sources (Sobolev 1992; Mehringer \& Menten 1996;
Salii, Sobolev \& Kalinina 2002), but are in contrast with the
generally held view of class I methanol masers derived from sources
such as DR 21(OH) (Plambeck \& Menten 1990) and G343.12-0.06
(Voronkov et al. 2006) where they are clearly associated with
outflow-molecular cloud interaction regions.

Our findings and those reported by Voronkov et al. (2010) are
inconsistent with some aspects of the evolutionary sequence
presented by Breen et al. (2010a) which has both the appearance and
disappearance of class I methanol masers preceding that of the class
II methanol masers. Breen et al. also have no overlap between the
class I methanol masers and OH maser stages. In our survey the EGOs
which are associated with class I methanol masers, but not class II
methanol masers are not found to be associated with any known
published H{\sc ii} regions or OH masers (actually there is absence
of any systemic surveys of H{\sc ii} regions or OH masers towards
EGOs at present). This suggests that these class I methanol masers
are likely to be excited by shocks driven from outflows, rather than
in the shocks driven by expanding H{\sc ii} regions at later
evolutionary stage. However, the true nature of these class I only
EGOs can only definitively be resolved by high resolution
observations which can determine the location of the maser emission
with respect to the EGOs. Because high-mass star formation regions
are crowded and frequently contain objects at a range of
evolutionary phases, chance associations are possible at low angular
resolutions. Examination of the results of Cyganowski et al. (2009)
for the source G28.28-0.36 illustrates how this can occur. For the
majority of the EGOs imaged by Cyganowski et al. (2009) in the 6.7
GHz class II and 44 GHz class I methanol maser transitions, both
types of masers are clearly associated with the targeted EGOs.
However, in the case of G28.28-0.36, while the 6.7 GHz class II
methanol masers are associated with the EGO, the class I masers are
offset and clearly trace the interface between an H{\sc ii} region
and the surrounding molecular gas. With single dish spatial
resolution both maser transitions and the EGO would be considered
likely to be coincident, although with the benefit of high
resolution data this is clearly not the case. This is only one
object from $\sim$20 observed by Cyganowski et al., so it is not
likely that all of the class I only EGO sources we have identified
are chance detections unrelated to the EGO, however, it is possible
that some may be.

The evolutionary scheme outlined by Ellingsen et al. (2007) and
Breen et al. (2010a) assume that each maser species arises only once
during the evolution of an individual massive star. However, it
appears that this assumption may require revision for class I
methanol masers. One possible manifestation of a two evolutionary
phase scenario for class I methanol masers (as discussed above)
would be that they initially arise at a relatively early phase of
the star formation process when powerful outflows interact with
surrounding molecular gas and that they are typically accompanied by
class II maser emission during this phase (the birthplace (disk or
outflow) of class II methanol masers remains uncertain, but at least
in some sources 6.7 GHz class II methanol masers can be excited in
the inner regions of outflows e.g. De Buizer 2003). As the source
evolves and the outflows diminish the class I maser emission fades
and ceases, but the class II maser emission continues, before it too
fades rapidly soon after the creation of the UC H{\sc ii} region. As
the ionized bubble rapidly expands it creates a second phase of
class I maser emission at the interface with the ambient molecular
gas. Of course, there is the possibility that the lifetime of class
I masers associated with outflows may also continue as far as the
stage when H{\sc ii} regions are detectable. Molecular line
observations of several massive star-forming regions show evidence
that outflows (and infall) can continue once ionization turns on,
and an UC H{\sc ii} region is formed (e.g. Keto \& Klaassen 2008,
Chen et al. 2010). However, high-mass star formation occurs in a
cluster environment which may include sequential or triggered star
formation, allowing a YSO associated with a well developed H{\sc ii}
region and a YSO associated with young outflow may coexist as near
neighbours. The extended spatial distribution of class I masers
compared to the other common maser species makes it more difficult
to determine which object the emission is associated with, outflow
or expanding H{\sc ii} region or both astrophysical phenomena
(particularly without high-resolution data). At present we cannot
determine whether the class I methanol masers associated with
outflows survive through to the stage when H{\sc ii} region appear,
although it appears possible from our results that this is the case.

At present we cannot confidently determine why the MIR colors of
class I only sources extends to less red colors than those
associated with class II masers.  It seems unlikely to be purely the
result of less reddening in these sources, but both association with
lower mass stars and their being more than one epoch of class I
maser emission remain plausible hypotheses.  Further observations an
millimetre and submillimetre wavelengths, combined with high
resolution observations of the class I masers will be required to
answer this question.

\subsection{The properties of mm dust clumps associated with methanol masers}

Table 1 shows that there are 63 EGOs in our observations which are
within the 1.1 mm continuum BGPS surveyed area (Rosolowsky et al.
2010). Among them, 54 are associated with a 1.1 mm BGPS sources,
while 9 are not (see also Table 2). A 1.1 mm BGPS source was
considered to be associated with an EGO if the separation between
the peak position of the BGPS source and the EGO position is less
than 30$''$. We did not take the size of the mm continuum source
into account when cross-matching (see Section 2.1).

In the two-evolutionary phase hypothesis for class I methanol masers
discussed in Section 4.2, we might expect similar trends to that
seen in the Mid-IR colors (e.g. Figure 6) to be present in other
physical tracers of the source evolution, such as the density of the
associated gas and dust. For example, Breen et al. (2010a; 2010b)
suggest that the density of the associated dust and gas decreases as
the sources evolve for class II methanol masers and water masers. So
in the two-evolutionary phase hypothesis we would predict class I
methanol only sources should have a lower gas density than class II
methanol maser only sources, which in turn would have a lower
density than those sources with both class I and II methanol masers.
To test this, we perform an investigation of the properties of 1.1
mm BGPS dust clumps associated with class I methanol masers in our
surveyed sample and Chen et al. (2009) sample (37 sources in total).
For each of the associated 1.1 mm BGPS source, we have assumed that
the 1.1 mm emission detected toward EGOs is from optically thin
dust. We can then calculate the gas mass using the equation:
\begin{equation}\label{1}
M_{gas}=\frac{S_{\nu}(int)D^{2}}{\kappa_{d}B_{\nu}(T_{dust})R_{d}},
\end{equation}
where $S_\nu$(int) is the 1.1 mm continuum integrated flux density,
$D$ is the distance to the source, $\kappa_{d}$ is the mass
absorption coefficient per unit mass of dust, $B_\nu(T_{dust})$ is
the Planck function for a blackbody at temperature $T_{dust}$, and
$R_d$ is the dust-to-gas mass ratio. Here we adopt $\kappa_{d}$$=$1
cm$^{2}$ g$^{-1}$ (Ossenkopf \& Henning 1994) for 1.1 mm emission,
and assume a dust-to-gas ratio ($R_{d}$) of 1:100. $B_\nu(T_{dust})$
was derived for an assumed dust temperature of 20 K. The H$_{2}$
column and volume densities of each dust clump were then derived
from its mass and radius (R$_{obj}$), assuming a spherical geometry
and a mean mass per particle of $\mu=2.29$ m$_{H}$, which takes into
account a 10\% contribution from helium (Fa\'{u}ndez et al. 2004).
We list the parameters of the 1.1 mm continuum integrated flux
density, $S_\nu$(int) and 1.1 mm source radius, R$_{obj}$ obtained
from the BGPS catalog (Rosolowsky et al. 2010) for all the 37 class
I maser sources with an associated 1.1 mm BGPS source in Table 7.
Dunham et al. (2010) suggested that a correction factor of 1.5 must
be applied to the Rosolowsky et al. BGPS catalog flux densities. In
this paper, we also apply this flux calibration correction factor to
the integrated flux density $S_\nu$(int) listed in Table 7. All the
associated 1.1 mm BGPS sources are resolved with the BGPS beam, with
the exception of G34.28+0.18. For this source we assumed half the
beam size (17$''$) as an upper limit for the object radius. Thus the
derived gas density of this source should be seen as a lower limit.
The masses and gas densities for all 1.1 mm dust clumps associated
with class I methanol masers determined using the methods outlined
above are listed in Table 7.

Figure 8 presents a log-log plot of the luminosity of the class I
methanol maser emission versus of the gas mass (left panel) and
H$_{2}$ density (right panel) of the associated 1.1 mm dust clump.
We have used different symbols in the plot to show whether the class
I masers are associated with class II methanol masers or not (21
with associated high-precision position class II masers; 9 without
high-precision position class II maser data; 7 without associated
class II masers; see Table 7). Figure 8 shows that a weak, but
statistically significant positive correlation exists in both cases.
We have performed a linear regression analysis for each
distribution, and plotted the relevant line of best fit in each
panel in Figure 8. The best fit linear equation for each
distribution is as follows:

log(L$_{m}$/L$_{sun}$)$=$0.50[0.14]log(M/M$_{sun}$)$-$6.34[0.435]

(correlation coefficient of 0.60 and p-value of 8.75e-04) and,

log(L$_{m}$/L$_{sun}$)$=$0.57[0.21]log(n(H$_2$))$-$7.25[0.75]

(correlation coefficient of 0.44 and p-value of 9e-03). These fits
demonstrate that the luminosity of the class I methanol masers
depends on the physical properties of the associated clump: the more
massive and denser the clump, the stronger the class I methanol
emission. We checked for correlations between the mass and density
of clump with the distance, but found no significant correlation in
either source property (mass or density) with distance.  This
suggests that the observed dependence between the 1.1 mm source
properties and class I maser luminosity is intrinsic and not an
observational artefact.

Interestingly, the dependence between the luminosity of class I
methanol masers and the gas density of the associated mm dust clump
is opposite to the relationship observed between the luminosity of
class II methanol masers and the gas density of mm dust clump
reported by Breen et al. (2010a). Breen et al. found the more
luminous 6.7 GHz class II methanol masers to be associated with mm
dust clumps with lower H$_{2}$ density, i.e. there is a negative
correlation between them. Even though their results were derived
with the peak luminosity rather than integrated luminosity of class
II methanol masers, the peak luminosity and integrated luminosity
are positively correlated. The simplest picture which fits the
different dependence between the luminosity of class I and class II
methanol masers with the gas density of the clump, is if the
intensity of class II methanol masers increases as the source
evolves/warms, while the class I maser intensity decreases as the
outflow broadens. If this is the case then we would expect to see an
anti-correlation in the class I/II flux densities. However,
comparing the peak flux density of class I and class II methanol
masers for our observed EGO sample, we have already shown that there
is no statistically significant correlation between them (see Figure
3).

We performed a similar analysis to that undertaken for the class I
masers associated with EGOs to check the relationship between the
peak luminosity of 6.7 GHz class II methanol maser and the gas
mass/density of the associated dust clump for the 21 EGOs associated
with both class I and II methanol masers (listed in Table 7). We
also found that there is a significant positive correlation between
the peak luminosity of the 6.7 GHz class II methanol maser and mass
of the associated dust clump (a slope of 1.21 with a standard error
of 0.32 and a p-value of 0.001 which allows us to reject the null
hypothesis of zero slope). The correlation coefficient between the
points was measured to be 0.66 for this linear regression analysis.
But there is no statistical correlation between the peak luminosity
of 6.7 GHz maser and the gas density (the linear regression analysis
shows a slope of -0.38 with a large p-value of 0.57 and a small
correlation coefficient of 0.15). The absence of correlation between
the peak luminosity of 6.7 GHz masers and gas density measured in
our analysis is not consistent with that of the anti-correlation
between these quantities found by Breen et al. (2010a). However the
sample size for this analysis is small and the class II masers are
clustered in a small range of parameter space with lower gas density
(log(n(H$_{2}$))$<$4), whereas the sample shown in Figure 2 of Breen
et al. covers a much wider range of gas densities. Thus our class II
maser sample is likely not representative of the larger population.
The correlation between the class I maser luminosity and gas density
is tighter than that measured for the 6.7 class II masers in our
analysis. The most likely explanation for this is that the gas
density of the clump is measured over a large spatial scale (the
angular resolution of the observations was typically 30$''$), which
can not accurately reflect the properties of the smaller compact
regions (of the order of 1$''$) associated with class II methanol
masers. In contrast the class I methanol maser spots usually have
much larger angular and spatial distributions (usually at the order
of 10$''$) and are associated with shocked regions (e.g. Cyganowski
et al. 2009). Thus we might expect that the gas densities derived
from dust clumps should correlate more closely with class I methanol
maser properties. The dependence of class I maser luminosity on
clump properties makes sense if the class I masers are excited at
the interface between outflows and surrounding material. Future
arcsecond resolution mm continuum imaging of these sources will be
necessary to determine if the relationship between the class II
maser properties and the associated gas and dust is tighter when the
spatial scales are comparable.

It is interesting to note that all of the sources associated with
only class I methanol masers are located in the bottom-left corner
of the right panel in Figure 8. This location corresponds to sources
with a lower maser luminosity and lower density of the dust clump.
The left panel of Figure 8 shows that median mass properties are
also different between the population associated with only class I
masers and that associated with both class I and II masers: the
population associated with both class I and II masers can extend to
higher clump masses than that associated with only class I masers.
The most widely accepted mechanism for massive star formation
suggests that high-mass sources are believed to originate from
massive clumps in the fragmentation of the giant molecular cloud.
The stellar mass of the sources is set by the fragmentation process
and the reservoir of material available to accrete is determined by
that as well (e.g. Hennebelle \& Chabrier 2008). Therefore the
measured distribution of 1.1 mm clump mass among different
populations indicates that these class II methanol maser EGO sources
may be associated with a higher stellar mass range than those where
only class I methanol masers are also observed. However, it is also
consistent with the predictions of the two-evolutionary phase
hypothesis for the class I only sources as the density of the
associated gas and dust decreases as the sources evolve (Breen et
al. 2010a; 2010b). So in the two-evolutionary phase hypothesis the
class I only sources (i.e. the later evolutionary sources) should
have a lower density than sources with only class II masers or with
both class I and II methanol masers (see the more detailed
discussions of this in Section 4.2). The assumed constant dust
temperature (T$_{dust}$) of 20 K in our calculations will affect the
results. However if the sources which are thought to be more evolved
(i.e. only class I maser sources) have a higher dust temperature, it
would result in a larger $B_\nu(T_{dust})$, thus a smaller mass and
lower density of dust clump associated with only class I masers. So
this is also credible in the two-evolutionary phase hypothesis.

However it must be noted that only 7 of the 31 class I maser-only
sources have estimates for the mass and density parameters from mm
dust continuum observations. The small size of this sample means
that it may not be representative of the entire population of class
I maser-only EGOs. Moreover some sources associated with both class
I and class II methanol masers also extend to the left-bottom corner
of each panel in Figure 8. This suggests that a small fractional of
the class II methanol maser population can also appear in an
environment of comparable mass and dust clump density to that seen
in the class I only associated EGOs. However, Figure 8 clearly shows
that the class II methanol masers are usually associated with more
massive and dense dust clumps than those associated with only class
I methanol masers.

\section{Conclusions}

Using the Mopra telescope, we have performed a systematic search for
95 GHz class I methanol masers toward EGOs. EGOs are new MYSO
candidates with ongoing outflows identified from the \emph{Spitzer}
GLIMPSE I survey. We detected 105 new 95 GHz masers from a sample of
192 targets. Of these, 92 have no previously observed class I
methanol maser activity, while the remaining 13 sources have been
detected in the 44 GHz transition. Thus our single-dish survey
proves that there is indeed a high detection rate ($\sim$55\%) of
class I methanol masers in EGOs. Our findings increase the number of
published class I methanol masers to 290 (an additional 92 on top of
the $\sim$198 from Val'tts \& Larinov 2010). Mid-IR color analysis
shows that the color-color region occupied by the GLIMPSE point
sources for EGOs which are and are not associated with class I
methanol masers are very similar, and mostly located in ranges
-0.6$<$[5.8]-[8.0]$<$1.4 and 0.5$<$[3.6]-[4.5]$<$4.0 (see section
4.1 for detailed discussion of the uncertainties involved in this
analysis). We find that the detection rate of class I methanol maser
is likely to be higher in those sources with redder GLIMPSE point
source colors.

Comparison of the [3.6]-[5.8] vs. [8.0]-[24] colors determined with
integrated fluxes from Cyganowski et al. (2008) for the subsamples
of the EGOs based on which class of methanol masers they are
associated with, shows that those which are only associated class I
methanol masers extend to less red colors than those associated with
both classes of methanol maser. We suggest that the less red colors
of class I methanol maser only EGOs is either because the class I
only EGOs are associated with lower stellar mass objects, or because
class I maser emission arises at more than one evolutionary phase of
the high-mass star formation process. On the basis of current
observations both scenarios can be plausibly argued and further
observations will be required to determine which, if either of these
hypotheses is correct. The thermal molecular line observations taken
in conjunction with our maser search will be useful for trying to
determine which of these scenarios is more likely. It will also be
important to undertake high resolution observations of the class I
maser emission in EGOs which are only associated with class I
methanol masers to determine where the maser emission arises
relative to the EGO. These observations are required to rule out the
possibility that these sources represent a sample of chance
associations between the EGOs and class I masers.

Analysis of the properties of mm dust clumps associated with class I
methanol masers (for a subset of the EGOs in the class I maser
survey sample which have available millimeter continuum data) shows
that the luminosity of the class I methanol masers is correlated
with the both the mass and density of the associated dust clump. The
more massive and denser the clump, the stronger the class I methanol
emission will be. We also find that the EGOs which are only
associated with class I methanol masers have a lower maser
luminosity and mass/density of dust clump. This finding supports
either the hypothesis that the class I maser can trace a population
with lower stellar masses, or that class I methanol masers may be
associated with more than one evolutionary phase during the
formation of a high-mass star.

\acknowledgements

We thank Dr. Karl Menten and an anonymous referee for their helpful
comments on this paper. We are grateful to the staff of the ATNF for
their assistance in the observation. The Mopra telescope is operated
through a collaborative arrangement between the University of New
South Wales and the CSIRO. This research has made use of the SIMBAD
database, operated at CDS, Strasbourg, France, and the data products
from the GLIMPSE survey, which is a legacy science program of the
{\em Spitzer Space Telescope}, funded by the National Aeronautics
and Space Administration. This work was supported in part by the
National Natural Science Foundation of China (grants 11073041,
10803017, 10625314, 10633010 and 10821302), the Knowledge Innovation
Program of the Chinese Academy of Sciences (Grant No. KJCX2-YW-T03),
the National Key Basic Research Development Program of China (No.
2007CB815405), and the CAS/SAFEA International Partnership Program
for Creative Research Teams.

\
\
\
\

\centerline {\textbf{REFERENCE}}

Batrla, W., \& Menten, K. M. 1988, ApJ, 329, L117

Becker, R. H., White, R. L., Helfand, D. J., \& Zoonematkermani, S.
1994, ApJS, 91, 347

Beuther, H., \& Steinacker, J. 2007, ApJ, 656, L85

Bohlin, R. C., Savage, B. D., \& Drake, J. F. 1978, ApJ, 224, 132

Bourke, T. L., Hyland, A. R., \& Robinson, G. 2005, ApJ, 625, 883

Breen, S. L., Ellingsen, S. P., Caswell, J. L., \& Lewis, B. E.
2010a, MNRAS, 401, 2219

Breen, S. L., Caswell, J. L., Ellingsen, S. P., \& Phillips, C. J.
2010b, MNRAS, 406, 1487

Carey, S. J., et al. 2009, PASP, 121, 76

Caswell, J. L. 1998, MNRAS, 297, 215

Caswell, J. L. 2009, PASA, 26, 454

Caswell, J. L., et al. 2010, MNRAS, 404, 1029

Chambers, E. T., Jackson, J. M., Rathborne, J. M., \& Simon, R.
2009, ApJS, 181, 360

Chen, X., Ellingsen, S. P., \& Shen, Z. Q. 2009, MNRAS, 396, 1603

Chen, X., Shen, Z. Q., Li, J. J., Xu, Y., \& He J. H. 2010, ApJ,
710, 150

Churchwell, E., et al. 2009, PASP, 121, 213

Cragg, D. M., Johns, K. P., Godfrey, P. D., \& Brown, R. D. 1992,
MNRAS, 259, 203

Cyganowski, C. J., et al. 2008, AJ, 136, 2391

Cyganowski, C. J., Brogan, C. L., Hunter, T. R., \& Churchwell, E.
2009, 702, 1615

Davis, C. J., Kumar, M. S. N., Sandell, G., Froebrich, D., Smith, M.
D., \& Currie, M. J. 2007, MNRAS, 374, 29

De Buizer, J. M., \& Vacca, W. D., 2010, AJ, 140, 196

De Buizer, J. M., 2003, MNRAS, 341, 277

Dunham, M. K., et al. 2010, ApJ, 717, 1157

Ellingsen, S.\ P. 2005, MNRAS, 359, 1498

Ellingsen, S. P. 2006, ApJ, 638, 241

Ellingsen, S. P. 2007, MNRAS, 377, 571

Ellingsen, S. P., Voronkov, M. A., Cragg, D. M., Sobolev, A. M.,
Breen, S. L., \& Godfrey, P. D. 2007, IAU Symp. 242, Astrophysical
Masers and their Environments, ed. J. M. Chapman \& W. A. Baan
(Cambridge: Cambridge Univ. Press), 213

Fa\'{u}ndez, S., Bronfman, L., Garay, G., Chini, R., Nyman,
L.-{\AA}., \& May, J. 2004, A\&A, 426, 97

Fazio, G. G., et al. 2004, ApJS, 154, 10

Forster, J. R., \& Caswell, J. L. 2000, ApJ, 530, 371

Green, J. A., et al. 2009, MNRAS, 392, 783

---, 2010, MNRAS, 409, 913

Gibb, A. G., \& Davis, C. J. 1998, MNRAS, 298, 644

Haschick, A. D., Menten, K. M., \& Baan, W. A. 1990, ApJ, 354, 556

Hennebelle, P., \& Chabrier, G. 2008, ApJ, 684, 395

Hill, T., Burton, M. G., Minier, V., Thompson, M. A., Walsh, A. J.,
Hunt-Cunningham, M., \& Garay, G. 2005, MNRAS, 363, 405

Indebetouw, R., et al. 2005, ApJ, 619, 931

Jackson, J. M., Finn, S. C., Rathborne, J. M., Chambers, E. T.,
Simon, R. 2008, ApJ, 680, 349

Johnston, K. J., Gaume, R. A., Wilson, T. L., Nguyen, H. A., \&
Nedoluha, G. E. 1997, ApJ, 490, 758

Kalenskii, S. V., Promyslov, V. G., Slysh, V. I., Bergman, P., \&
Winnberg A. 2006, Astron. Rep., 50, 289

Kalenskii, S. V., Johansson, L. E. B., Bergman, P., Kurtz, S.,
Hofner, P., Walmsley, C. M., \& Slysh, V. I. 2010, MNRAS, 405, 613

Keto, E. \& Klaassen, P. 2008, ApJ, 678, L109

Kurtz, S., Churchwell, E., \& Wood, D. O. S. 1994, ApJS, 91, 659

Kurtz, S., Hofner, P., \& \'Alvarez, C. V. 2004, ApJS, 155, 149

Ladd, N., Purcell, C.,Wong, T., \& Robertson, S. 2005, Publ. Astron.
Soc. Aust., 22, 62

Mehringer, D. M., \& Menten, K. M. 1996, ApJ, 474, 346 Menten K., M.
1991a, in ``Skylines'' proceedings of the Third Haystack Observatory
Meeting, ed. A.\ D.\ Haschick, P.\ T.\ P.\ Ho (San Fransisco:
Astronomical Society of the Pacific), 119

Menten, K. M. 1991b, ApJ, 380, L75

Minier, V., Ellingsen, S. P., Norris, R. P., \& Booth, R. S. 2003,
A\&A, 403, 1095

Noriega-Crespo, A., et al. 2004, ApJS, 154, 352

Ossenkopf, V., \& Henning, T. 1994, A\&A, 291, 943

Pandian, J. D., Goldsmith, P. F., \& Deshpande, A. A. 2007, ApJ,
656, 255

Pandian, J. D., Leurini, S., Menten, K. M., Belloche, A., \&
Goldsmith, P. 2008, A\&A, 489, 1175

Peretto, N., \& Fuller, G. A. 2009, A\&A, 505, 405

Pestalozzi, M. R., Minier, V., \& Booth, R. S. 2005, A\&A, 432, 737

Pillai, T., Wyrowski, F., Carey, S. J., \& Menten, K. M. 2006, A\&A,
450, 569

Plambeck, R. L., \& Menten, K. M. 1990, ApJ, 364, 555


Povich, M. S., \& Whitney, B. A. 2010, ApJ, 714, 285

Povich, M. S., et al. 2009, ApJ, 696, 1278

Ragan, S. E., Bergin, E. A., \& Gutermuth, R. A. 2009, ApJ, 698, 324

Rathborne, J. M., Jackson, J. M., Chambers, E. T., Stojimirovic, I.,
Simon, R., Shipman, R., \& Frieswijk, W. 2010, ApJ, 715, 310

Reach, W. T., et al. 2006, AJ, 131, 1479

Reid, M. J., et al. 2009, ApJ, 700, 137

Rohlfs, K., \& Wilson, T. L. (ed.) 2004, Tools of Radio Astronomy
(4th revised and enlarged edition; Berlin: Springer)

Robitaille, T. P., Whitney, B. A., Indebetouw, R., Wood, K., \&
Denzmore, P. 2006, ApJS, 167, 256

Robitaille, T. P., et al. 2008, AJ, 136, 2413

Rosolowsky, E., et al. 2010, ApJS, 188, 123

Salii, S. V., Sobolev, A. M., \& Kalinina, N. D. 2002, Astron. Rep.,
46, 955

Smith, H. A., Hora, J. L., Marengo, M., \& Pipher, J. L. 2006, ApJ,
645, 1264

Slysh, V. I., Kalenskii, S. V., Val'tts, I. E., \& Otrupcek, R.
1994, MNRAS, 268, 464

Sobolev, A. M. 1992, SvA, 36, 590


Val'tts, I. E., Ellingsen, S. P., Slysh, V. I., Kalenskii, S. V.,
Otrupcek, R., \& Larinov, G. M. 2000, MNRAS, 317, 315

Val'tts I.\ E., \& Larinov G.\ M., 2007, Astron. Rep., 51, 519

Val'tts, I. E., Larionov, G. M., \& Bayandina, O. S. 2010, eprint
arXiv:1005.3715


Voronkov, M. A., Brooks, K. J., Sobolev, A. M., Ellingsen, S. P.,
Ostrovskii, A. B., \& Caswell J.\ L. 2006, MNRAS, 373, 411

Voronkov, M. A., Caswell, J. L., Ellingsen, S. P., \& Sobolev, A.
M., 2010, MNRAS, 405, 2471

Walsh, A. J., Burton, M. G., Hyland, A. R., \& Robinson, G. 1998,
MNRAS, 301, 640

Whitney, B. A., et al. 2004, ApJS, 154, 315

Wood, D. O. S., \& Churchwell, E. 1989, ApJS, 69, 831

Xu, Y., Li, J. J., Hachisuka, K., Pandian, J. D., Menten, K. M., \&
Henkel, C. 2008, A\&A, 485, 729

Xu, Y., Voronkov, M. A., Pandian, J. D., Li, J. J., Sobolev, A. M.,
Brunthaler, A., Ritter, B., \& Menten, K. M. 2009, A\&A, 507, 1117

Ybarra, J. E., Lada, E. A., Balog, Z., Fleming, S. W., \& Phelps, R.
L. 2010, ApJ, 714, 469

Ybarra, J. E., \& Lada, E. A. 2009, ApJ, 695, 120


\tabletypesize{\scriptsize}

\setlength{\tabcolsep}{0.08in}

\begin{deluxetable}{llcccccccccccc}
\tablewidth{0pt} \tablecaption{Sample parameters}
\renewcommand{\arraystretch}{1.0}
\tablehead{ &\multicolumn{2}{c}{Position$^a$}& \\
\cline{2-3}
Source & R.A. (2000)& Dec. (2000)& IRDC$^{b}$ & Class II$^c$ & OH$^d$ &  UC H{\sc ii}$^d$ & mm $^d$ & Remark$^e$ & class I$^f$\\
 & ($\emph{h m s}$) & (\arcdeg \ \arcmin \ \arcsec) &  & }

 \startdata

  G10.29$-$0.13$^l$                                     &   18  08  49.3    &   $-$20   05  57  &   Y     &  Y     &    --     &      N    &   Y    &    2   &   Y     \\
  G10.34$-$0.14$^l$                                     &   18  09  00.0    &   $-$20   03  35  &   Y     &  Y     &    --     &      N    &   Y    &    2   &   Y     \\
  G11.11$-$0.11$^{i^\ast}$$^{m^\dagger}$                &   18  10  28.3    &   $-$19   22  31  &   Y     &  Y     &    --     &      N    &   Y    &    3   &   N     \\
  G11.92$-$0.61$^g$$^{h^\ast}$$^{l^\dagger}$            &   18  13  58.1    &   $-$18   54  17  &   Y     &  Y     &    --     &      --   &   --   &    1   &   Y     \\
  G12.02$-$0.21$^i$$^k$                                 &   18  12  40.4    &   $-$18   37  11  &   Y     &  N     &    --     &      N    &   Y    &    1   &   N     \\
  G12.20$-$0.03$^j$$^l$                                 &   18  12  23.6    &   $-$18   22  54  &   N     &  Y     &    --     &      Y    &   Y    &    4   &   Y     \\
  G12.42$+$0.50$^n$                                     &   18  10  51.1    &   $-$17   55  50  &   N     &  N     &    --     &      --   &   Y    &    4   &   Y     \\
  G12.68$-$0.18$^l$                                     &   18  13  54.7    &   $-$18   01  47  &   N     &  Y     &    Y      &      N    &   Y    &    4   &   Y     \\
\enddata
\end{deluxetable}

\clearpage


\begin{deluxetable}{lccccccc}
\tabletypesize{\tiny}
 \setlength{\tabcolsep}{0.01in}

\tablewidth{0pt} \tablecaption{Summary of the other datasets utilised in this work} \tablehead{Category & Datasets & (Un)Targeted &
Area covered & Angular resolution & Sensitivity$^a$ & Comments$^b$}
 \startdata

\textbf{OH maser }            & Caswell (1998)             & Untargeted &  $|b|<1^\circ$ \&        &  $\sim$ 7$''$ &  $\sim$ 0.12 Jy                 &  101 EGOs fell within the area  \\
                              &                            &            &  $l=312^\circ-356^\circ$ &               &                                 &   11 detected                    \\
                              &                            & Targeted   &  $l=230^\circ-13^\circ$  &  $\sim$ 7$''$ &  $\sim$ 0.12 Jy                 &  additional 3 EGOs were detected            \\
                              \\
\textbf{Class II maser}       & Caswell (2009)             & Targeted   &   $l=188^\circ-50^\circ$ &  $\sim$ 2$''$ &  ...                            &  $\sim$ 60 EGOs were detected   \\
                              & Cyganowski et al. (2009)   & Targeted   &   ...                    &  $\sim$ 2$''$ &  $\sim$ 0.1 Jy                  &  $\sim$ 20 EGOs were detected         \\
                              & Xu et al. (2009)           & Targeted   &   ...                    & a few arcsecs &  ...                            &  $\sim$ 30 EGOs were detected        \\
                              & Caswell et al. (2010)      & Untargeted &   $|b|<2^\circ$ \&       & a few arcsecs &  $\sim$ 0.2 Jy                  &  11 EGOs fell within the area,  \\
\enddata

\tablenotetext{a} {The sensitivities quoted here are uniformly set
to 3 $\sigma_{rms}$ for all surveys. For the Class II maser surveys
of Caswell et al. (2010) and Green et al. (2010), we quoted the
sensitivity from the subsequent ATCA observation for the Parkes MMB
survey.}

\tablenotetext{b} {Comments on the extent of overlap with our Morpa
surveyed EGOs, except for class II maser datasets that also include
51 EGOs listed in our previous work (Chen et al. 2009) but which
were not observed in our Mopra survey (see Table 6).}

\end{deluxetable}


\tabletypesize{\scriptsize}

\setlength{\tabcolsep}{0.12in}
\renewcommand{\arraystretch}{0.98}

\begin{deluxetable}{lccccccccc}

\tablewidth{0pt} \tablecaption{Observed properties of 95 GHz class I
methanol maser sources detected.}
\tablehead{  &  \multicolumn{4}{c}{Gaussian Fit} & \\
\cline{2-5}
Source & $S$ & V$_{LSR}$ &  $\Delta V$ & $P$ & $\sigma_{rms}$ & S$_{int}$ & D & L$_{m}$ \\
 & (Jy km s$^{-1}$) & (km s$^{-1}$) & (km s$^{-1}$) & (Jy) & (Jy)  & (Jy km s$^{-1}$)  &  (kpc) & (10$^{-6}$ L$_{\odot}$) \\
 (1) & (2) & (3) & (4) & (5) & (6)& (7) & (8) & (9)}
\startdata

G10.29-0.13     &    10.43(0.92)   &     13.96(0.41)    &    8.41(0.88)    &   0.99   &    0.55   &  12.87   &   2.3   &  2.1    \\
                &     1.62(0.24)   &     14.31(0.03)    &    0.46(0.07)    &   2.82   &           &          &         &         \\
                &     0.82(0.18)   &     15.13(0.03)    &    0.28(0.07)    &   2.30   &           &          &         &         \\
G10.34-0.14     &    39.05(0.81)   &     12.24(0.05)    &    5.30(0.12)    &   6.92   &    0.62   &  69.21   &   2.3   &  11.5   \\
                &    30.16(0.26)   &     14.74(0.00)    &    0.45(0.00)    &   63.05  &           &          &         &         \\
G11.92-0.61     &    22.55(0.33)   &     33.94(0.01)    &    0.82(0.01)    &   25.70  &    0.62   &  69.51   &   3.8   &  31.5   \\
                &    28.63(0.45)   &     35.10(0.00)    &    0.52(0.01)    &   52.18  &           &          &         &         \\
                &    18.34(0.60)   &     36.26(0.02)    &    1.44(0.06)    &   11.95  &           &          &         &         \\
\enddata
\tablecomments{Column (1): source name. Columns (2)-(5): the
integrated flux density S, the velocity at peak maser emission
V$_{LSR}$, the line width (FWHM) $\Delta$V, and the peak flux
density P, of each maser feature estimated from Gaussian fits to 95
GHz class I methanol maser lines, the formal error from the Gaussian
fit is given in parenthesis. The corresponding values in main beam
temperature T$_{MB}$ (K) can be obtained by dividing the flux
density by a factor of 9.3 (see Section 2.2). Column (6): 1$\sigma$
noise in the observed maser spectrum. Column (7): the total
integrated flux density S$_{int}$ of the maser spectrum obtained
from summing the integrated flux density of all maser features in
each source in column (2). Column (8): the kinematic distance to
source estimated from galactic rotation curve of Reid et al. (2009).
The distances to G49.07-0.33, G309.91+0.32, G310.15+0.76 which
cannot be derived from the galactic rotation curve are assumed to be
5 kpc. Column (9): the integrated luminosity of 95 GHz methanol
maser estimated with assuming maser isotropic emission, i.e.
L$_{m}$=4$\pi$$\cdot$D$^2$$\cdot$S$_{int}$.
\\
(This table is available in its entirety in a machine-readable form
in the online journal. A portion is shown here for guidance
regarding its form and content.)}

\end{deluxetable}
\clearpage


\tabletypesize{\scriptsize}

\setlength{\tabcolsep}{0.05in}


\begin{deluxetable}{lclclclc}

\tablewidth{0pt} \tablecaption{Sources undetected at 95 GHz.}
\tablehead{ Source & $\sigma_{rms}$ (Jy) & Source & $\sigma_{rms}$
(Jy) & Source & $\sigma_{rms}$ (Jy) & Source & $\sigma_{rms}$ (Jy)}
\startdata

G11.11-0.11     &    0.71   &  G12.02-0.21     &    0.61   &  G16.58-0.08     &    0.56   & G17.96+0.08     &    0.60     \\
G19.61-0.12     &    0.45   &  G20.24+0.07     &    0.63   &  G21.24+0.19     &    0.63   & G24.17-0.02     &    0.52     \\
G28.28-0.36     &    0.48   &  G28.85-0.23     &    0.45   &  G29.89-0.77     &    0.45   & G29.91-0.81     &    0.52     \\
G29.96-0.79     &    0.56   &  G35.15+0.80     &    0.45   &  G35.68-0.18     &    0.56   & G36.01-0.20     &    0.63     \\
G37.48-0.10     &    0.60   &  G37.55+0.20     &    0.56   &  G40.28-0.27     &    0.48   & G44.01-0.03     &    0.47     \\
G45.47+0.05     &    0.48   &  G45.50+0.12     &    0.48   &  G48.66-0.30     &    0.47   & G49.42+0.33     &    0.50     \\
G53.92-0.07     &    0.58   &  G54.11-0.04     &    0.74   &  G54.11-0.08     &    0.60   & G57.61+0.02     &    0.74     \\
G298.90+0.36    &    0.58   &  G304.89+0.64    &    0.61   &  G305.57-0.34    &    0.43   & G305.62-0.34    &    0.65     \\
\enddata
\end{deluxetable}


\begin{deluxetable}{lccc}
\tabletypesize{\normalsize}
 \setlength{\tabcolsep}{0.20in}

\tablewidth{0pt} \tablecaption{Detection rates of class I methanol
maser in different subsamples} \tablehead{Source Properties &
N$_{D}$$^a$ & N$_{T}$$^a$ & Detection rate }
 \startdata

Likely outflow                          &  53  &  86  & 62\%   \\
Possible outflow                        &  52  & 106  & 49\%   \\
IRDC                                    &  71  & 128  & 55\%   \\
Non-IRDC                                &  34  & 64   & 53\%   \\
With associated class II maser          &  39  & 49   & 80\%   \\
Without associated class II maser       &  31  & 81   & 38\%   \\
With associated OH maser                &  13  & 14   & 93\%   \\
Without associated OH maser             &  43  & 90   & 48\%   \\
With associated UC H{\sc ii} region     &  11  & 13   & 85\%   \\
Without associated UC H{\sc ii} region  &  21  & 34   & 62\%   \\
With associated 1.1 mm                  &  35  & 54   & 65\%   \\
Without associated 1.1 mm               &   1  &  9   & 11\%   \\

\enddata
\tablenotetext{a} {N$_{D}$ and N$_{T}$ represent the numbers of the
detected 95 GHz class I maser sources and total sources,
respectively.}

\end{deluxetable}


\tabletypesize{\scriptsize}

\setlength{\tabcolsep}{0.03in}

\begin{deluxetable}{lccclccclccc}
\tablewidth{0pt} \tablecaption{Sources used in Chen et
al. (2009) analysis but not observed in the present Mopra survey}
\renewcommand{\arraystretch}{1.0}
\tablehead{ Source & Class II$^a$ &  Class I$^b$ & Remark$^c$ &
Source & Class II$^a$ & Class I$^b$ & Remark$^c$ }

\startdata

G14.33-0.64$^d$$^{h^\dagger}$                    &    Y   &   Y    &  1   &     G331.34-0.35$^f$$^h$                      &    Y   &   Y   &   4   \\
G16.59-0.05$^{f^\ast}$$^h$                       &    Y   &   Y    &  2   &     G332.29-0.09$^h$                          &    Y   &   Y   &   4   \\
G23.01-0.41$^{d^\ast}$$^{h^\dagger}$             &    Y   &   Y    &  1   &     G332.35-0.44$^i$                          &    Y   &   N   &   4   \\
G28.83-0.25$^e$$^{i^\dagger}$                    &    Y   &   N    &  1   &     G332.56-0.15$^e$$^{i^\dagger}$            &    Y   &   N   &   1   \\
G43.04-0.45                                      &    Y   &   --   &  4   &     G332.60-0.17$^h$                          &    Y   &   Y   &   2   \\
G298.26+0.74                                     &    Y   &   --   &  1   &     G332.73-0.62                              &    Y   &   --  &   2   \\
G309.38-0.13$^{d^\ast}$$^{h^\dagger}$            &    Y   &   Y    &  1   &     G332.94-0.69$^d$$^{h^\dagger}$            &    Y   &   Y   &   1   \\
G318.05+0.09$^f$$^i$                             &    Y   &   N    &  2   &     G332.96-0.68$^d$$^{f^\ast}$$^{h^\dagger}$ &    Y   &   Y   &   1   \\
\enddata
\end{deluxetable}


\begin{deluxetable}{lccccccccc}
\tabletypesize{\scriptsize}
 \setlength{\tabcolsep}{0.05in}
\renewcommand{\arraystretch}{1.0}
\tablewidth{0pt} \tablecaption{Properties of mm dust clumps
associated with methanol masers}

\tablehead{EGO name & BGPS name$^a$ & R$_{obj}$$b$ & S$_{v}$(int)$^b$ & M$^c$ & n(H$_{2}$)$^c$ & N(H$_{2}$)$^d$ & A$_{v}$$^d$ \\
  & & ($''$) & (Jy) & (M$_{\odot}$) & (10$^3$ cm$^{-3}$) &  (10$^{22}$ cm$^{-2}$) }

\startdata

G10.29$-$0.13$^h$                 &   G010.286$-$00.120   &      68  &   9.41    &  1100   &    10.5       &   3.3    &  33   \\
G10.34$-$0.14$^h$                 &   G010.343$-$00.144   &      85  &   8.58    &  1000   &    5.0        &   1.9    &  19   \\
G12.20$-$0.03$^f$$^h$             &   G012.201$-$00.034   &      53  &   2.06    &   940   &    3.2        &   1.2    &  12   \\
G12.42$+$0.50$^j$                 &   G012.419+00.506     &      51  &   0.76    &    99   &    6.8        &   0.5    &  5   \\
G12.68$-$0.18$^h$                 &   G012.681$-$00.182   &      86  &   11.82   &  5900   &    5.6        &   2.6    &  26   \\
G12.91$-$0.03$^e$$^h$$^\dagger$   &   G012.905$-$00.030   &      65  &   3.56    &  1800   &    2.2        &   1.4    &  14   \\
G12.91$-$0.26$^h$                 &   G012.909$-$00.260   &      79  &   15.85   &  5200   &    1.7        &   4.1    &  41   \\
G14.63$-$0.58$^f$$^h$$^\dagger$   &   G014.633$-$00.574   &      72  &   10.20   &  1100   &    5.4        &   3.2    &  32   \\
\enddata

\tablenotetext{a}{The associated 1.1 mm BGPS continuum source
identified from BGPS catalog (Rosolowsky et al. 2010).}

\tablenotetext{b}{The associated 1.1 mm BGPS source radius and flux
density used in the calculation. Note that a flux calibration
correction factor of 1.5 was needed to apply to the flux density
listed here when calculating the gas mass (see Dunham et al. 2010).
All sources are resolved with the BGPS beam expect G34.28+0.18. We
assumed the beam size as an upper limit on the object radius of this
source.}

\tablenotetext{c}{The gas mass and beam-averaged volume density of
clump derived from the corresponding 1.1 mm continuum source. The
beam size of BGPS is 33$''$.}

\tablenotetext{d}{The beam-averaged column density N(H$_{2}$) and
reddening vector A$_{v}$ of clump. We adopt A$_{v}=10^{-21} \times$
N(H$_{2}$) from Bohlin et al. (1978).}

\tablenotetext{e, f} {The sources associated with GLIMPSE point
sources whose colors locate outside and in the color region occupied
by Stage I derived by Robitaille et al. (2006) in Figure 5,
respectively. See more details in Section 4.1.}

\tablenotetext{g}{The sources are selected from Chen et al. (2009)
sample.}

\tablenotetext{h, i, j} {Represent the three different subsamples: h
is a source associated with both a class I methanol maser and a
high-precision postion class II methanol maser (including 21
members), i is a source associated with class I masers but without
high-precision position class II maser information (including 9
members), and h is a source associated with only class I maser
emission and without class II maser detection by Mt Pleasant
(including 7 members).}

\tablenotetext{\dagger}{Sources with integrated flux measurements
from Cyganowski et al. (2008) are also overlapped in the [3.6]-[5.8]
vs. [8.0]-[24] color analysis of Figure 6.}

\end{deluxetable}

\begin{figure*}
\caption{Overlays of the OH and 6.7 GHz class II masers, UC H{\sc
ii} regions, and 1.1 mm BGPS sources on the \emph{Spitzer} 3-color
IRAC images with 8.0 $\mu$m (red), 4.5 $\mu$m (green) and 3.6 $\mu$m
(blue) for all 192 targeted EGOs. The yellow contours are the 24
$\mu$m MIPSGAL data (Carey et al. 2009) (the contour levels for each
source are not presented). The positions of OH masers, 6.7 GHz class
II methanol masers, UC H{\sc ii} regions, and 1.1 mm BGPS sources
are denoted by small red circles, black crosses, blue squares and
yellow diamonds, respectively. The targeted point is marked by a
blue plus. The large white circle represents the region covered by
the Mopra beam, with a solid circle for detected and dashed circle
for undetected 95 GHz class I methanol masers, respectively. (A
color and complete version of this figure is available in the online
journal.)}
\end{figure*}

\begin{figure*}
\scalebox{1}[1]{\includegraphics[55,-10][500,500]{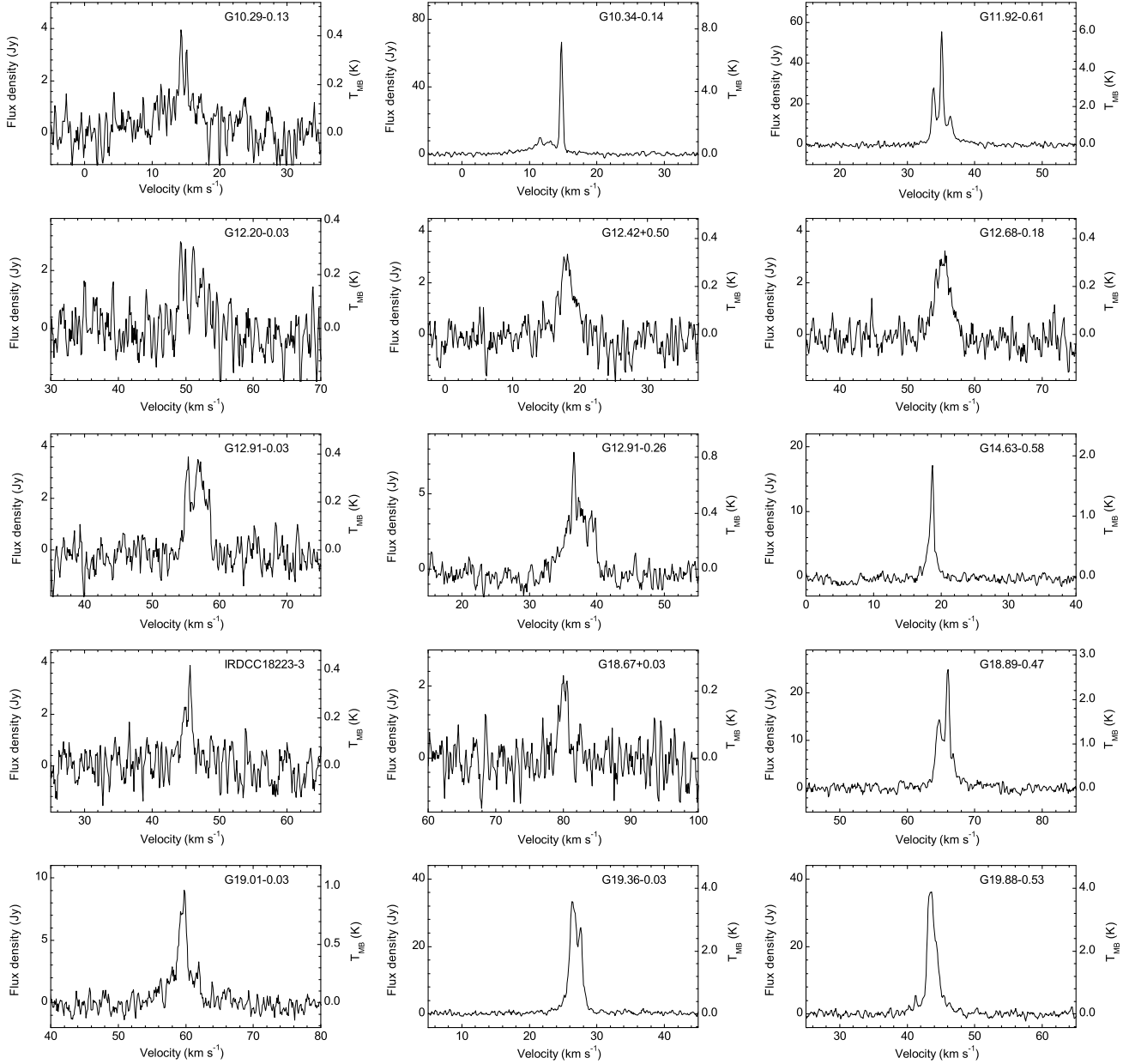}}
\caption{Spectra of the 95 GHz methanol maser sources detected in
the EGO-based searches. The left and right labels of Y-axis show the
values in flux density and main beam temperature, respectively for
each panel. Note that the Y-axis scale is not the same
panel-to-panel. The velocity range covering 40 km s$^{-1}$ shown in
X-axis is chosen to locate the emission approximately in the middle
for each panel.}
\end{figure*}

\begin{figure*}
\scalebox{1}[1]{\includegraphics[55,0][500,500]{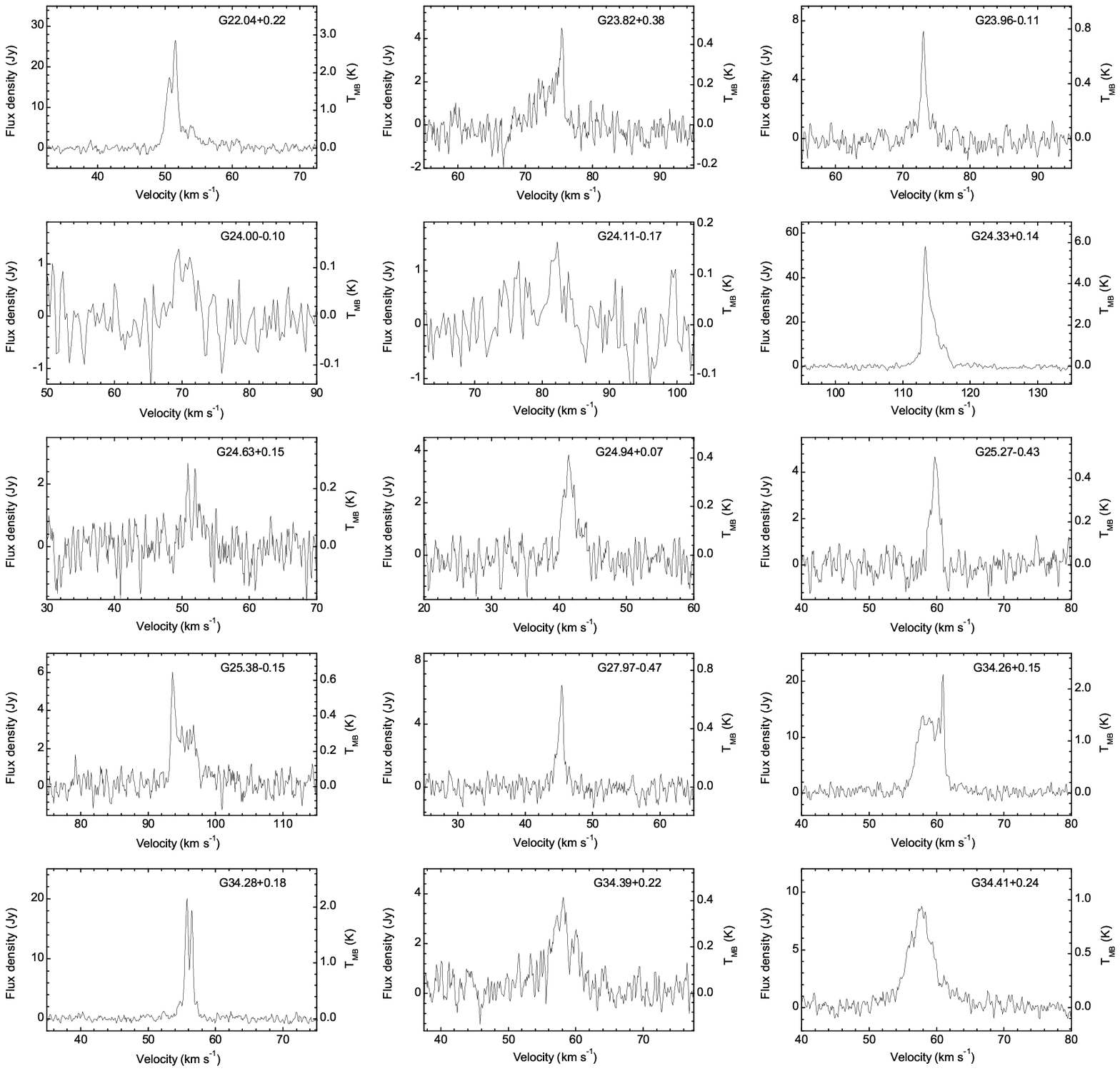}}
\vspace{6mm}

Fig. 2.--- Continued.
\end{figure*}

\begin{figure*}
\scalebox{1}[1]{\includegraphics[55,0][500,500]{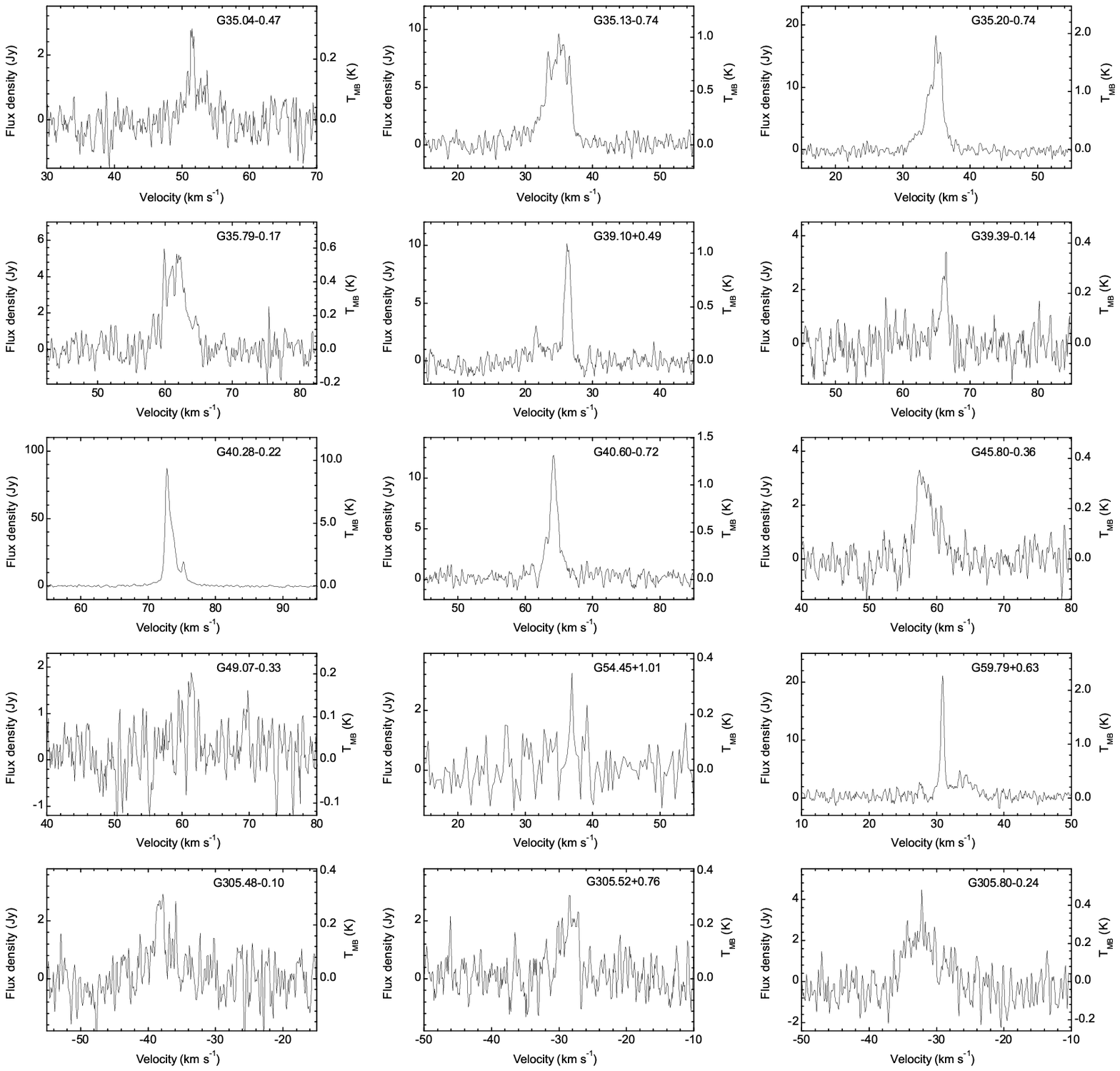}}
\vspace{6mm}

Fig. 2.--- Continued.
\end{figure*}

\begin{figure*}
\scalebox{1}[1]{\includegraphics[55,0][500,500]{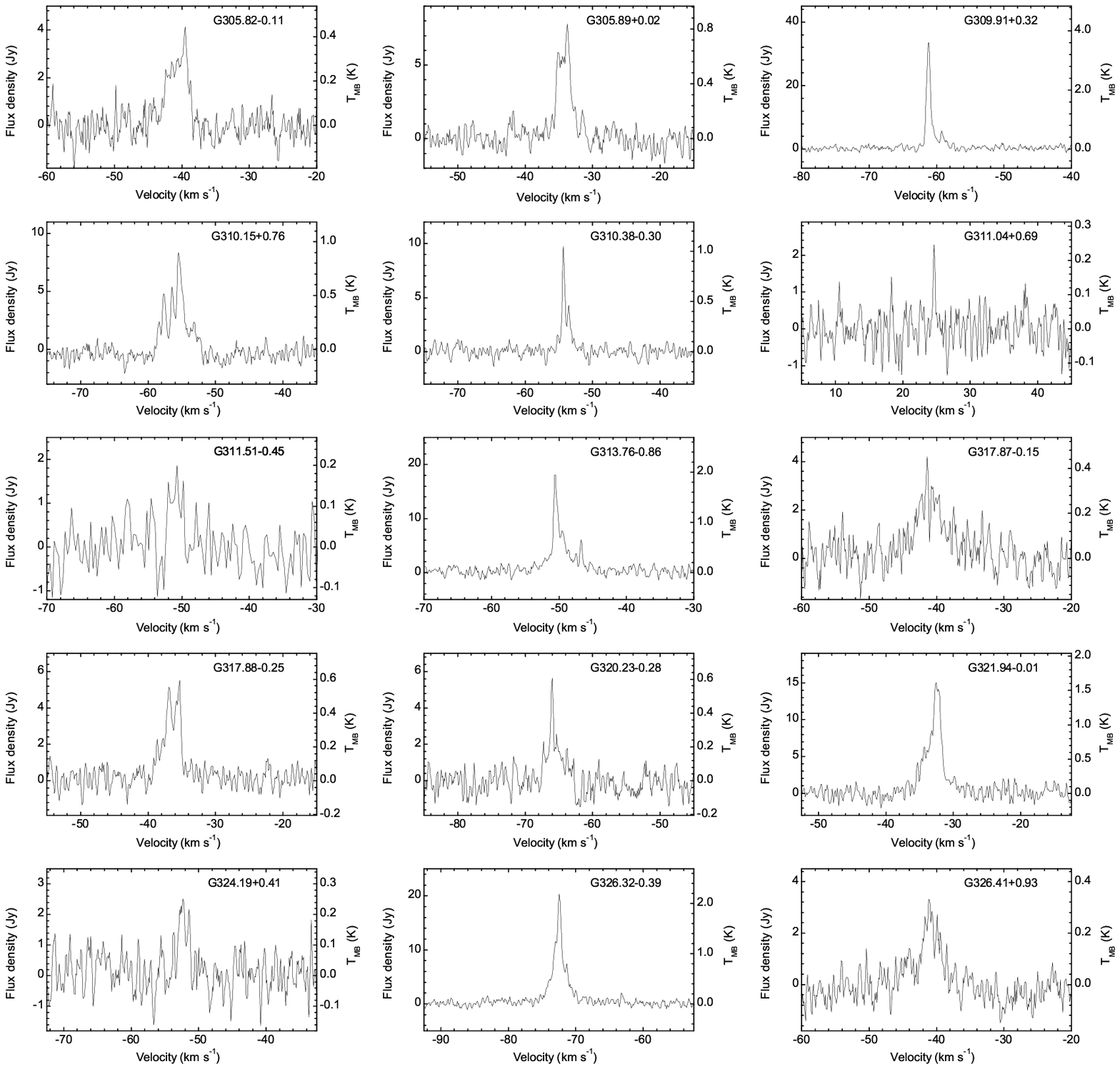}}
\vspace{6mm}

Fig. 2.--- Continued.
\end{figure*}

\begin{figure*}
\scalebox{1}[1]{\includegraphics[55,0][500,500]{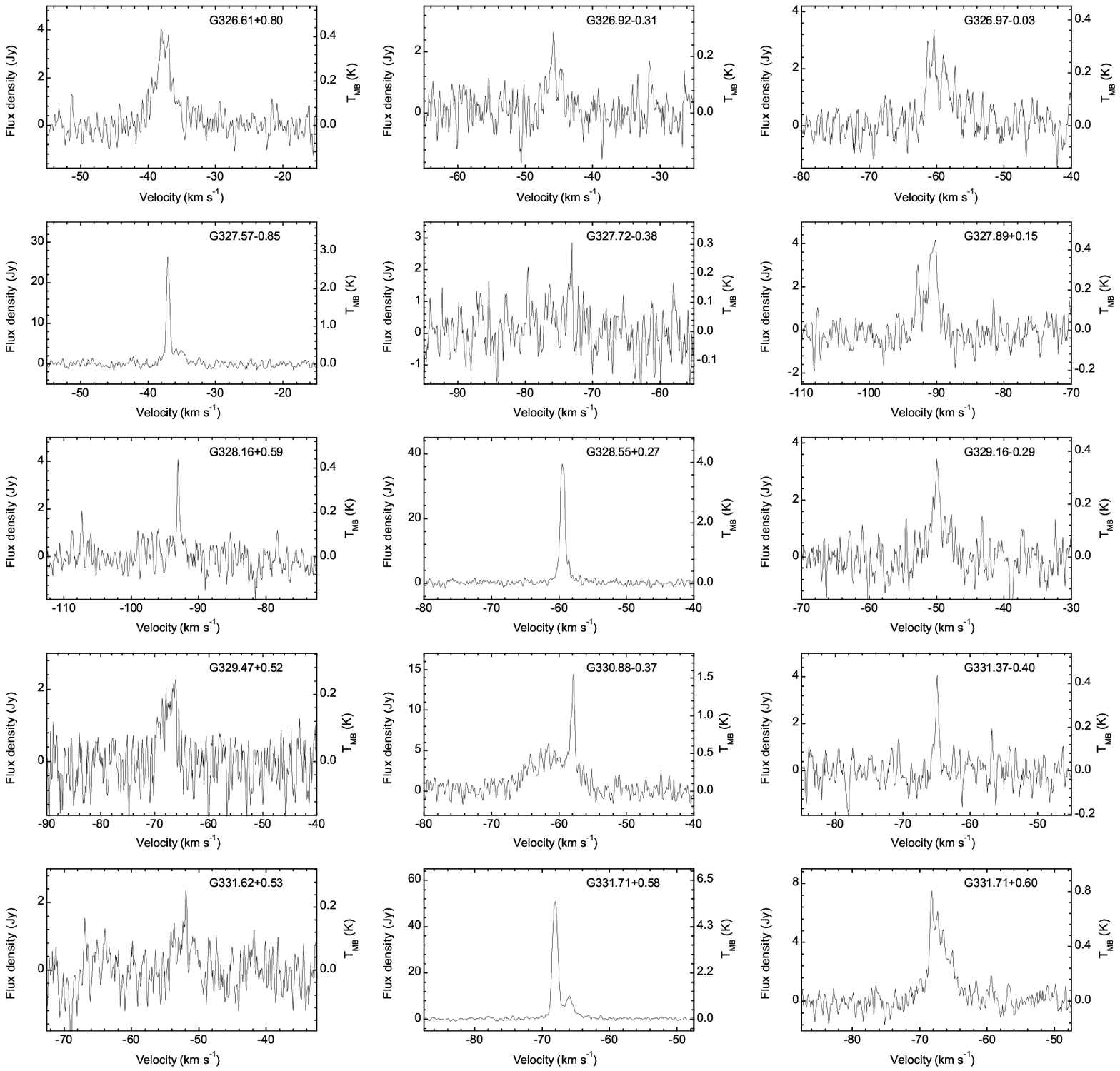}}
\vspace{6mm}

Fig. 2.--- Continued.
\end{figure*}

\begin{figure*}
\scalebox{1}[1]{\includegraphics[55,0][500,500]{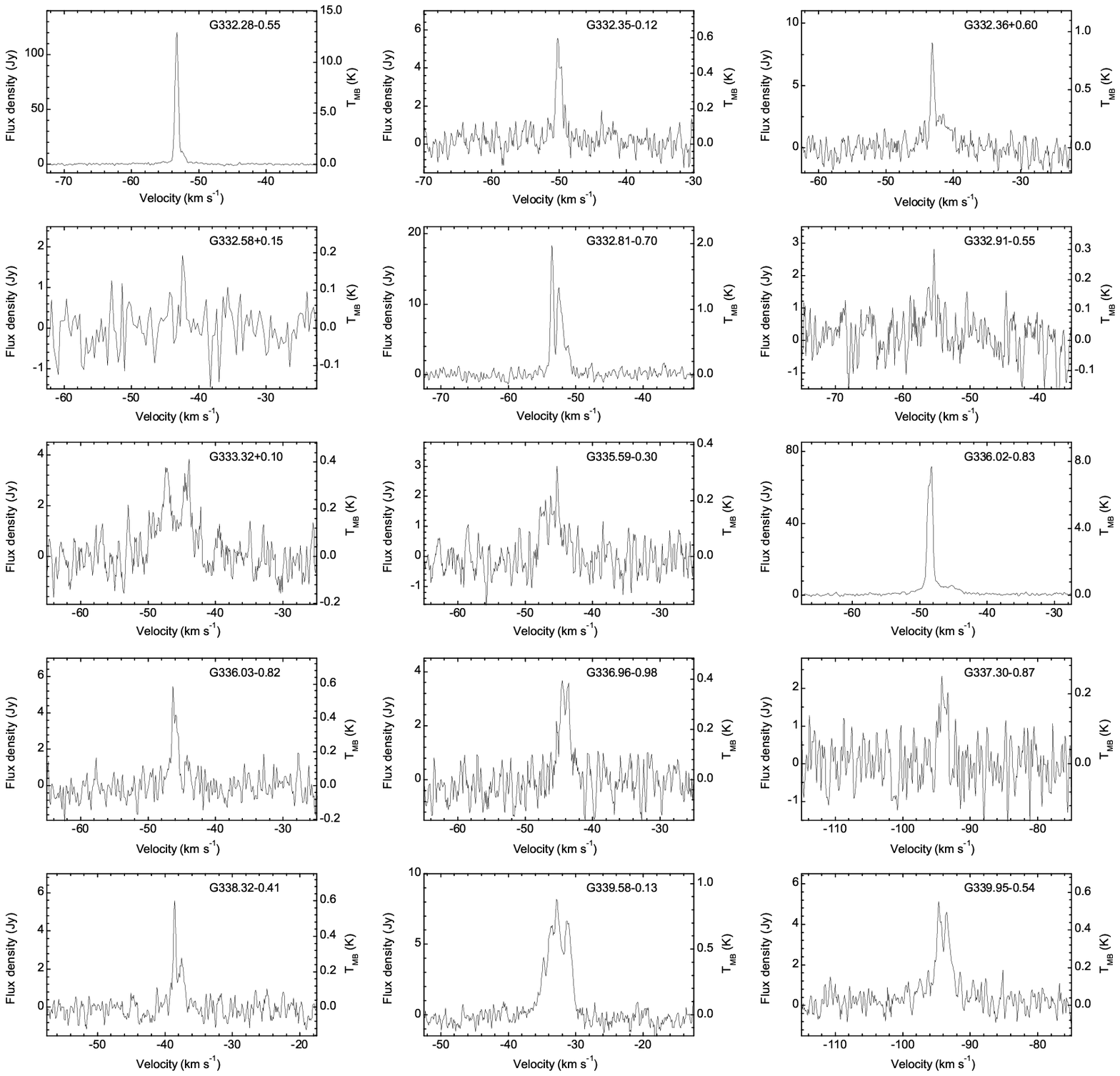}}
\vspace{6mm}

Fig. 2.--- Continued.
\end{figure*}

\begin{figure*}
\scalebox{1}[1]{\includegraphics[55,0][500,500]{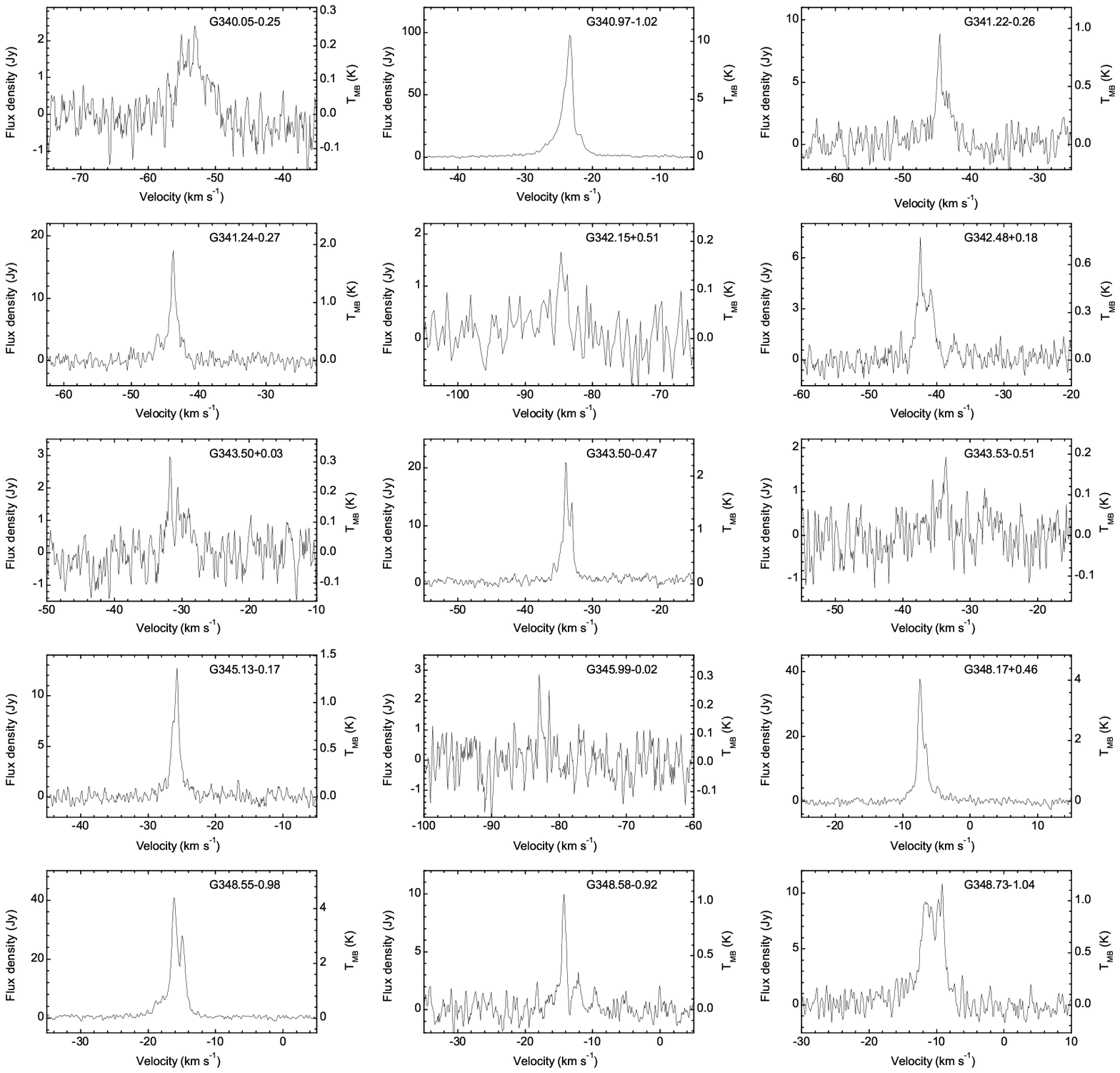}}
\vspace{6mm}

Fig. 2.--- Continued.
\end{figure*}

\begin{figure*}
\scalebox{0.7}[0.7]{\includegraphics[-30,0][300,600]{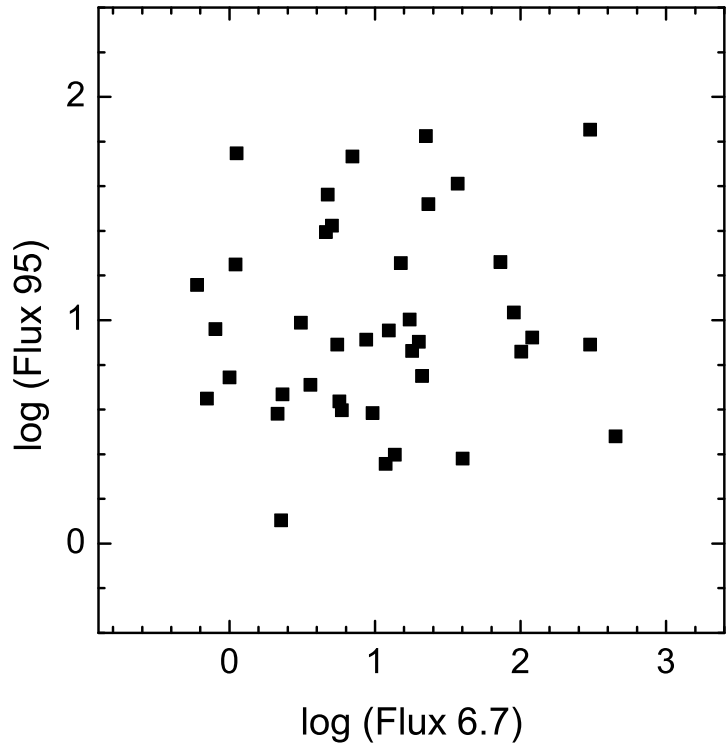}}
\caption{Comparison of the distribution of the peak flux density of
the 95 GHz class I methanol masers (represented by Flux 95) to that
of the 6.7 GHz class II methanol masers (represented by Flux 6.7)
for the EGOs associated with both classes of masers in our observing
sample.}

\end{figure*}

\begin{figure*}
\scalebox{0.9}[0.9]{\includegraphics[-30,00][700,500]{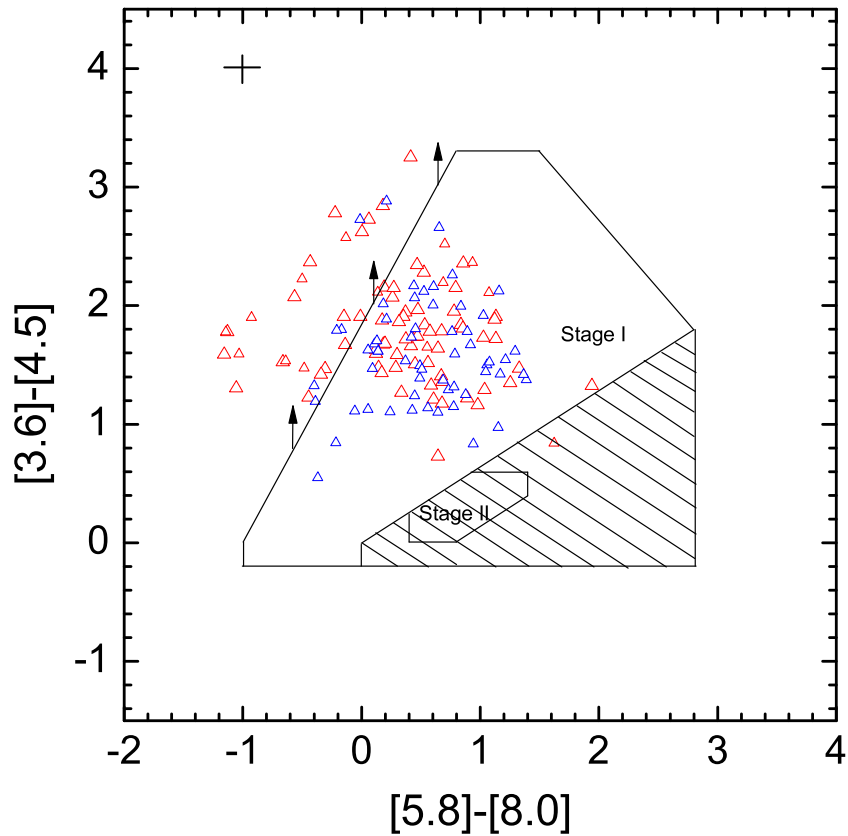}}
\caption{[3.6]-[4.5] vs. [5.8]-[8.0] color-color plot of EGOs. Only
sources listed in Table 1 and 3 of Cyganowski et al. (2008) for
which there is flux density measurements for all the four IRAC bands
are plotted. The red and blue triangles represent the EGOs which are
and are not associated with class I methanol masers, respectively.
The solid lines mark the regions occupied by various
evolutionary-stage (Stages I, II and III) YSOs according to the
models of Robitaille et al. (2006). The hashed region in the
color-color plot are regions where models of all evolutionary stages
can be present. The error bar in the top left was derived from the
average standard deviation of the measurements of all data in the
plot. The reddening vectors show an extinction of A$_{v}$=20,
assuming the Indebetouw et al. (2005) extinction law.}
\end{figure*}

\begin{figure*}
\scalebox{0.8}[0.8]{\includegraphics[-30,0][700,500]{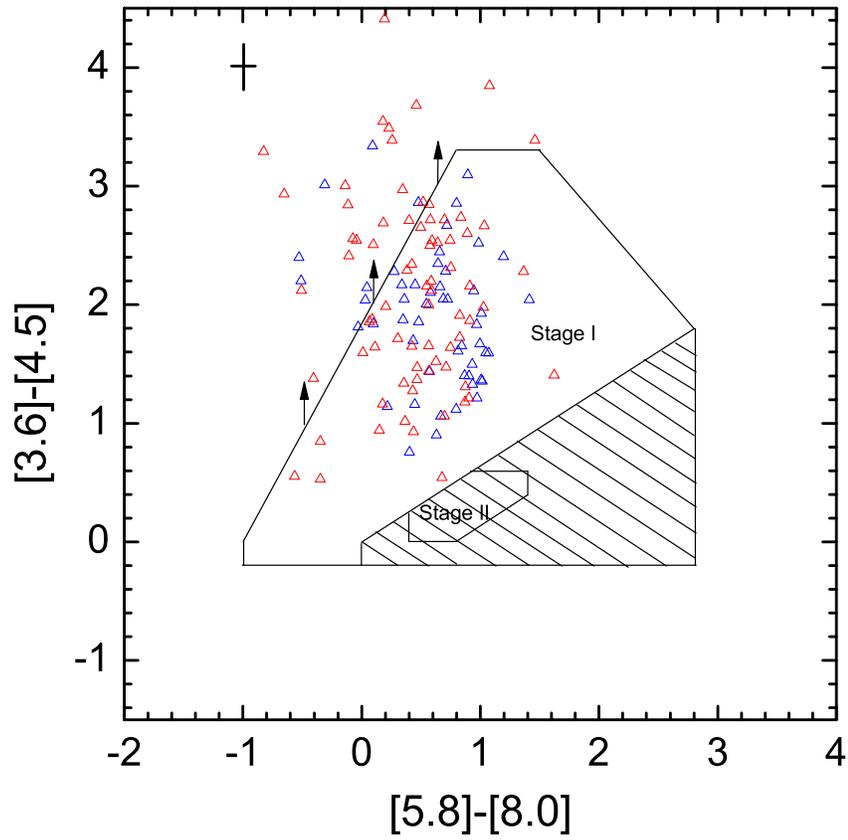}}
\caption{Same as Figure 4, but for the nearest GLIMPSE point sources
associated with EGOs. Only sources for which there is flux density
measurements for all the four IRAC bands are plotted.}
\end{figure*}

\begin{figure*}
\scalebox{0.83}[0.83]{\includegraphics[-30,0][400,400]{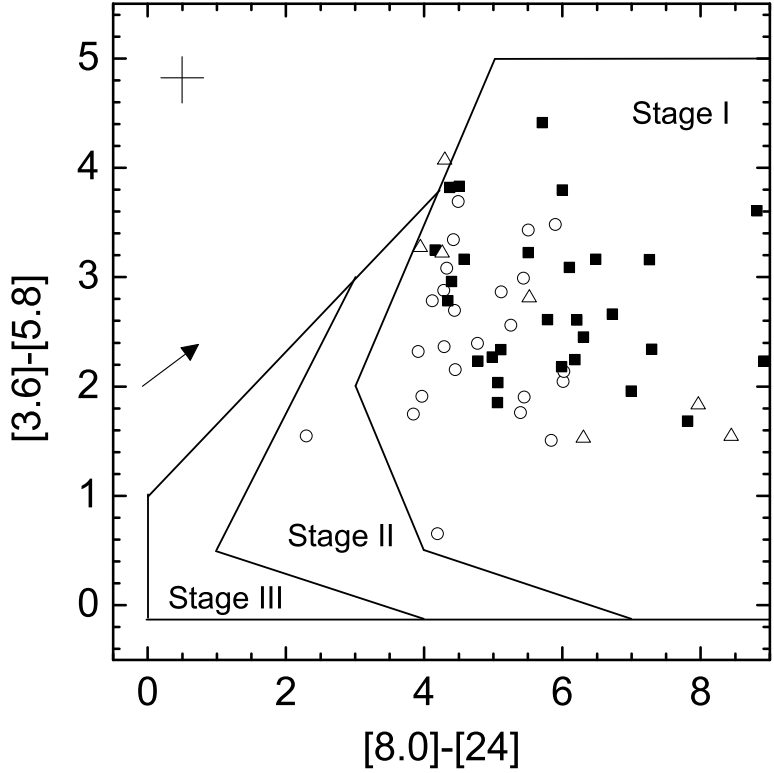}}
\caption{[3.6]-[5.8] vs. [8.0]-[24] color diagram of EGOs associated
with three subsamples based on which of class methanol masers they
are associated with (see Section 4.2): associated only with class I
methanol masers (marked by open circles), associated only with known
6.7 GHz class II methanol masers with high accurate positions
(marked by open triangles), and associated with both class I and
high accurate position 6.7 GHz class II methanol masers (marked by
filled squares). Only sources listed in Table 1 and 3 of Cyganowski
et al. (2008) for which there is flux density measurements for all
the four IRAC bands and MIPS 24 $\mu$m are plotted. The error bar in
the top left of each plot was derived from the average standard
deviation of the measurements of all data in the corresponding plot.
The arrow in each plot represents a reddening vector at an
extinction of A$_{v}$=20 derived from the Indebetouw et al. (2005)
extinction law.}
\end{figure*}

\begin{figure*}
\scalebox{0.8}[0.8]{\includegraphics[00,10][300,400]{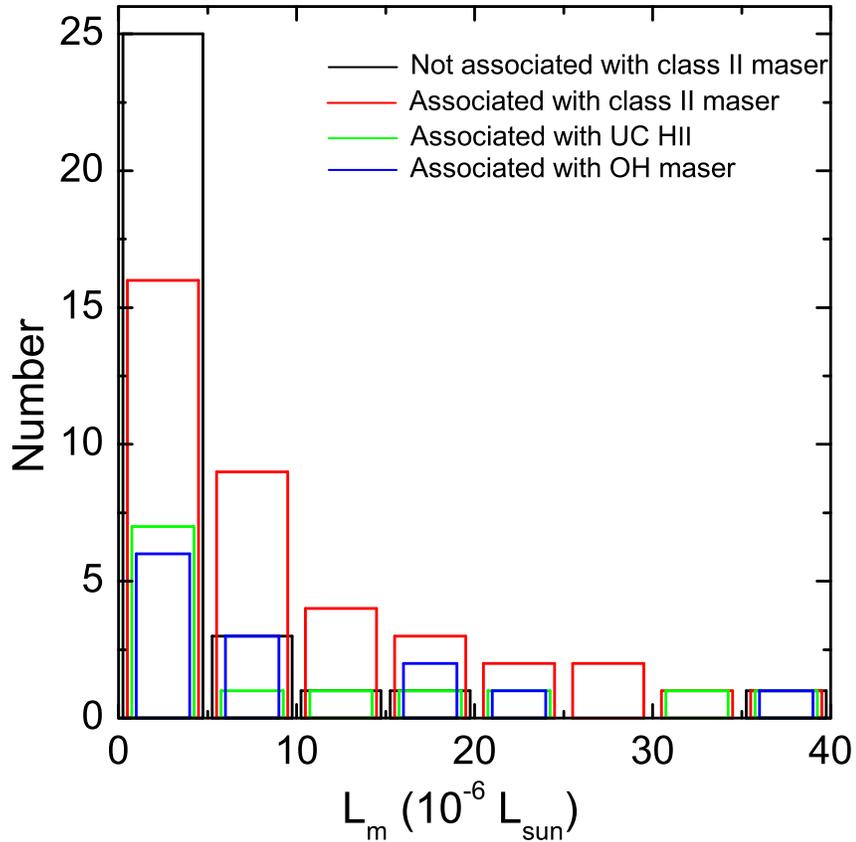}}
\caption{Histogram of the luminosity of 95 GHz class I methanol
maser detected in our observations with various associations. The
different color bins represent the different associations marked in
the right-top corner. The class II methanol maser subsample includes
the Mopra-surveyed EGOs (39 in total) associated with high-precision
position class II masers within 30$''$ (see Table 1 and Section
4.2).}
\end{figure*}
\clearpage

\begin{figure*}
\scalebox{0.9}[0.9]{\includegraphics[40,90][400,400]{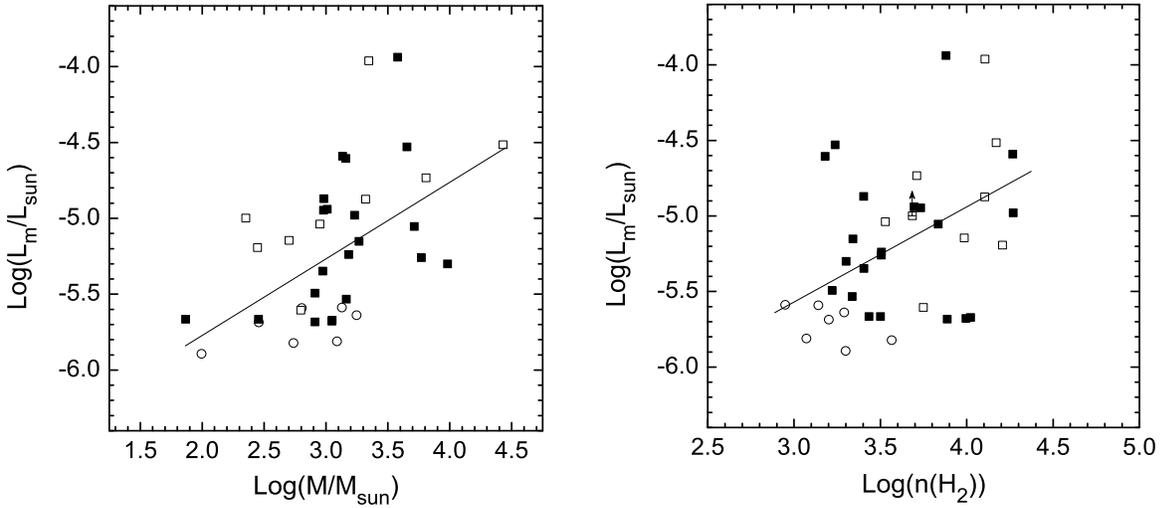}}
\caption{Logarithm of the 95 GHz class I methanol maser luminosity
as a function of the gas mass (left panel) and H$_{2}$ density
(right panel) of the associated 1.1 mm dust clump. The filled
squares, open squares and open circles represent the class I maser
sources which are with high-precision position class II methanol
maser associated (21 members), without high-precision position class
II maser information (9 members), and without an class II maser
detection by Mt Pleasant (7 members), respectively. The line in each
panel marks the best fit to the corresponding distribution. The
upward arrow in the right-hand panel indicates the lower limit for
the gas density of the source G34.28+0.18.}
\end{figure*}

\end{document}